\documentclass[showpacs,a4paper,nofootinbib,superscriptaddress]{revtex4-2}
\usepackage[utf8]{inputenc}
\usepackage{float}
\usepackage{graphicx}
\usepackage{subcaption}
\usepackage{epsfig}
\graphicspath{ {./images/} }

\usepackage{newpxmath,newpxtext}
\usepackage[export]{adjustbox}
\usepackage{graphicx}
\usepackage{epsfig}
\usepackage[normalem]{ulem}
\usepackage{xcolor}
\usepackage[a4paper]{geometry}
\usepackage{tensor}
\usepackage{verbatim}

\newcommand{\pder}[2]{\dfrac{\partial#1}{\partial#2}}
\newcommand{\pdder}[3]{\dfrac{\partial^2 #1}{\partial #2 \partial #3}}

\newcommand{\dder}[2]{\dfrac{\delta#1}{\delta#2}}
\newcommand{\pdot}[1]{\dot{\partial}_{#1}}

\newcommand{\Gd}{\mathcal{G}}

\newcommand{\R}{\mathcal{R}}
\newcommand{\de}{\mathrm{d}}

\newcommand{\lin}{\\[7pt]}

\title{Schwarzschild-Finsler-Randers spacetime:\\
Dynamical analysis, Geodesics and Deflection Angle}

\begin{document}
\title{Schwarzschild-Finsler-Randers spacetime:\\Geodesics, Dynamical Analysis
and Deflection Angle}
% Author 1
\author{E.~Kapsabelis}
\email{manoliskapsabelis@yahoo.gr}
\affiliation{Section of Astrophysics, Astronomy and Mechanics, Department of 
Physics, National and Kapodistrian University of Athens, Panepistimiopolis 15784, Athens, Greece}

% Author 2
\author{P.~G.~Kevrekidis}
\email{	kevrekid@umass.edu}
\affiliation{Department of Mathematics and Statistics, University of Massachusetts, Amherst, MA 01003-9305, USA}

% Author 3
\author{P.~C.~Stavrinos}
\email{pstavrin@math.uoa.gr}
\affiliation{Department of Mathematics, National and Kapodistrian University of 
Athens,	Panepistimiopolis 15784, Athens, Greece}

% Author 4
\author{A.~Triantafyllopoulos}
\email{alktrian@phys.uoa.gr}
\affiliation{Section of Astrophysics, Astronomy and Mechanics, Department of 
Physics, National and Kapodistrian University of Athens, Panepistimiopolis 15784, Athens, Greece}

\begin{abstract}
In this work, we extend the study of Schwarzschild-Finsler-Randers (SFR) spacetime 
previously investigated by a subset of the present authors \cite{Triantafyllopoulos:2020vkx, Kapsabelis:2021dpb}.
%which we investigated in two previous works .  
We will examine the dynamical analysis of geodesics which provides the derivation of the energy and the angular momentum of a particle moving along a geodesic of SFR spacetime. This study allows us to compare our model with the corresponding of general relativity (GR).
In addition, the effective potential of SFR model is examined and it is compared with the effective potential of GR. The phase portraits generated
by these effective potentials are also compared.
Finally we deal with the derivation of the deflection angle of the SFR spacetime and we find that there is a small perturbation from the deflection angle of GR. %which could perhaps be attributed to Lorentz violations. 
%{\bf Is it perhaps a good idea not to mention this in the abstract?
% I.e., stick to the results here and avoid the speculation?}
We also derive an interesting relation between the deflection angles of SFR model and the corresponding
result
in the work of~\cite{Shapiro}.
%Shapiro's estimated one.}  
% This was not only done by Shapiro...
These small differences are attributed to the anisotropic metric structure of the model and especially 
to a Randers term which provides a small deviation from the GR.
\end{abstract}
\maketitle

\section{Introduction}\label{sec: Introduction}

Einstein’s field equations in general relativity predict that the curvature is produced not only by the distribution of mass-energy but also by its motion \cite{Hartle2002}. Candidate metric geometries that can intrinsically describe the motion are the Finsler and Finsler-like geometries which constitute metrical generalizations of Riemannian geometry and depend on position and velocity/momentum/scalar coordinates. These are dynamic geometries that can describe locally anisotropic phenomena and Lorentz violations
\cite{Asanov:1991rmp, Stavrinos:1999rmp, Kostelecky:2011qz, AlanKostelecky:2012yjr, Stavrinos:2012kv, Minguzzi:2015xka, Foster:2015yta, Antonelli:2018fbv, Edwards:2018lsn, Ikeda:2019ckp, Relancio:2020mpa}
as well as with field equations, FRW and Raychaudhuri equations, geodesics, dark matter and dark energy effects \cite{Konitopoulos:2021eav, Stavrinos:2021ygh, Kouretsis:2008ha, Mavromatos:2010jt, Stavrinos:2020llm, Hama:2022vob}.
By considering this approach, the gravitational field is interpreted as the metric of a generalized spacetime and constitutes a force-field which contains the motion. This possibility  reveals the Finslerian geometrical character of spacetime.

In the framework of applications of Finsler geometry, many works in different directions of geometrical and physical structures have contributed to the extension of research for theoretical and observational approaches during the last years. We cite some works from the literature of the applications of Finsler geometry
\cite{Gibbons:2007iu, Skakala:2010hw, Kostelecky:2011qz, AlanKostelecky:2012yjr, GallegoTorrome:2012qra, Stavrinos:2013neo, Fuster:2015tua, Voicu:2015uta, Hohmann:2018rpp, Edwards:2018lsn, Colladay:2019lig, Caponio:2020ofw, Hama:2021frk, Hama:2022vob, Li:2022rjv}.

In the first period of development of applications of Finsler geometry to Physics, especially to General Relativity, remarkable works were 
published by G.~Randers~\cite{Randers1941}, J.~I.~Horv{\'a}th~\cite{Horvath1950} and A.~Mo{\'o}r~\cite{HorvathMoor}.   
Later, Einstein’s field equations were formulated in the Finslerian framework by the works of J.~I.~Horv{\'a}th~\cite{Horvath1950,HorvathMoor},
Y.~Takano~\cite{Takano1968} and S.~Ikeda~\cite{Ikeda1981}. In these studies, the field equations had been considered without calculus of variations.     
G.~S.~Asanov~\cite{Asanov1983} explored the Finslerian gravitational field by using Riemannian osculating methods and derived Einstein field equations using the variational principle.
%\\[7pt]
A class of Finsler spaces (FR standing
for Finsler-Randers) originated by 
G.~Randers~\cite{Randers1941} who studied the physical properties of spacetime with an asymmetrical metric which provides the uni-direction of time-like intervals. This consideration gives a particular interest in a generalized metric structure of the Riemannian spacetime.
%The type of FR space constitutes a significant kind of a Finsler space. 
% {\bf This should be explained or better off
% removed}
Based on this form of  spacetime, it is possible to investigate the gravitational field with more degrees of freedom in the framework of a tangent/vector/scalar bundle \cite{Triantafyllopoulos:2018bli, Triantafyllopoulos:2020ogl, Hama:2021frk}. 
The FR cosmological model was first introduced in \cite{stavrinos2005, Stavrinos:2006rf}. It is of special interest since the Friedmann equations include an extra geometrical term that acts as a dark energy-fluid.
The Finsler-Randers-type spacetime can be considered as a direction-dependent motion of the Riemannian/FRW model.

The local anisotropic structure of spacetime affects the gravitational field and leads to modified cosmological considerations. Based on Finsler or Finsler-like cosmologies, the Friedmann equations include extra terms which influence the cosmological evolution \cite{Kouretsis:2008ha, Triantafyllopoulos:2018bli,Hama:2021frk, Konitopoulos:2021eav}.
When Lorentz symmetry holds, the spacetime is isotropic in the sense that all directions and uniform motions are equivalent. 
The introduction of a vector field in the structure of spacetime causes relativity violations and local anisotropy which arise from breaking the Lorentz symmetry and which affect the metric, curvature, geodesics and null cone \cite{Girelli:2006fw, Minguzzi:2014aua, Javaloyes:2018lex, Silva:2013xba, Kostelecky:2008be, Vacaru:2010fi, Stavrinos:2012ty, Hohmann:2016pyt}.

In the framework of modified gravitational theories with Finsler-Randers type structure, two fundamental theories of investigation for the gravitation and cosmology can be developed. The first one is connected with the Friedmann-Finsler-Randers cosmological model and the second one is related to the study of SFR spacetime.

The introduction of a force field causes an asymmetry to a pseudo-Finslerian metric. Asymmetrical and locally anisotropic models such as Finsler-Randers spacetime can be connected with the chiral fields in Cosmology for descriptions of the inflationary epoch and the present accelerated expansion of the Universe \cite{Chervon:2013btx}.

An FR space has a metric function of the form
	\begin{equation}\label{lagrangian}
		F(x,y) = (-a_{\mu\nu}(x)y^{\mu}y^{\nu})^{1/2} + u_{\alpha}y^{\alpha}    
	\end{equation}
	where $u_{\alpha}$ is a covector with $||u_{\alpha}||\ll 1$, $y^{\alpha}=\frac{dx^{\alpha}}{d\tau}$ and $a_{\mu\nu}(x)$ is a pseudo-Riemannian metric for which the Lorentzian signature $(-,+,+,+)$ has been assumed and the indices $\mu, \nu, \alpha$ take the values $0,1,2,3$. The geodesics of this space can be produced by \eqref{lagrangian} and the Euler-Lagrange equations. If $u_{\alpha}$ denotes a force field $f_{\alpha}$ and $y^{\alpha}$ is substituted with $d x^{\alpha}$ then $f_{\alpha}dx^{\alpha}$ represents the spacetime effective energy produced by the anisotropic force field $f_{\alpha}$, therefore equation \eqref{lagrangian} is written as
	\begin{equation}\label{lagrangian2}
		F(x,dx) = \left(-a_{\mu\nu}(x)dx^{\mu}dx^{\nu}\right)^{1/2} + f_{\alpha}dx^{\alpha}    
	\end{equation}
 This form of metric provides a dynamical effective structure of spacetime. A small differentiation is presented between  GR and the FR gravitation model. This is because of the work provided by the one-form $A_{\gamma}$ which gives an external motion  to the Riemannian spacetime. This motion is an internal concept for the FR spacetime.\\
 
A cosmological model can be introduced by Eq.~(\ref{lagrangian2}) if we assume the FRW cosmological metric instead of the general type of the Riemannian one \cite{stavrinos2005, Stavrinos:2006rf}. In this case, we get a Friedmann-Finsler-Randers cosmological model in the following form 	
	\begin{equation}
		a_{\mu\nu}(x) = \mathrm{diag}\left[-1,\frac{a^2}{1-\kappa r^{2}},a^{2}r^{2}, a^{2}r^{2}\sin^{2}\theta\right]   
	\end{equation}
This model was also further studied later in \cite{Stavrinos:2002rc, Chang:2007vq, Stavrinos:2012kv, Basilakos:2013hua, Basilakos:2013ij, Brody:2015zra, Silva:2015ptj, Stavrinos:2016xyg, Papagiannopoulos:2017whb, Chaubey:2018wph, Chanda:2019guf, Chanda:2019mro, Heefer:2020hra, Raushan:2020mkh, Papagiannopoulos:2020mmm, Silva:2020tqr, Lou:2021gwk, Hama:2022vob, Angit:2022lfu}.

In this work, we will follow the SFR model presented in \cite{Triantafyllopoulos:2020vkx}. The metric $g_{\mu\nu}$ is the classic Schwarzschild one:
	\begin{align}\label{Schwarzchild}
		g_{\mu\nu}\de x^\mu  \de x^\nu 
		 = -fdt^2 + \frac{dr^2}{f} + r^2 d\theta^2 + r^2 \sin^{2}\theta\, d\phi^2
	\end{align}
 with $f=1-\frac{R_s}{r}$ and $R_s=2GM$ the Schwarzschild radius (we assume units where the speed of light $c=1$).
	
Hereafter, we consider an $\alpha$-Randers type metric as the one in Eq.~\eqref{lagrangian} which is distinguished from the $\beta$-Randers type metric that is investigated in the Standard Model Extension (SME) \cite{Kostelecky:2011qz,AlanKostelecky:2012yjr,Silva:2013xba,Foster:2015yta}.
	
The metric $v_{\alpha\beta}$ is derived from a metric function $F_v$ of the $\alpha$-Randers type \cite{Triantafyllopoulos:2020vkx}:
	\begin{equation}\label{RandersL}
		F_v = \sqrt{-g_{\alpha\beta}(x)y^\alpha y^\beta} + A_\gamma(x) y^\gamma
	\end{equation}
where $g_{\alpha\beta}=g_{\mu\nu}\tilde\delta^{\mu}_{\alpha}\tilde\delta^{\nu}_{\beta}$ is the Schwarzschild metric from Eq.~\eqref{Schwarzchild} and $A_{\gamma}(x)$ is a covector which expresses a deviation from general relativity, with $|A_\gamma(x)|\ll 1$, i.e., 
we assume that the deviation is small.
In this work, we continue the investigation of the Schwarzschild-Finsler-Randers spacetime (SFR) which  has been studied in previous works
 by a subset of the present authors \cite{Triantafyllopoulos:2020vkx, Kapsabelis:2021dpb}.
 
 In this article we examine the influence of extra gravitational effect which is imprinted in the geodesics of an SFR spacetime and we compare it with that of GR case. We prove that the additional amount of energy included in the equation of geodesics (see Eqs.~\eqref{geodesics0}-\eqref{geodesics3}) stems from the geometry and the form of SFR model.
 
 This approach is also applied the Newtonian gravitational theory and we obtain an interesting relation between the fundamental term $A_{0}(r)$ of SFR gravitational theory and the Newtonian potential. This relation also provides a physical interpretation of the extra term $A_{0}(r)$ in the Newtonian framework.
 
 In addition we calculate the deflection angle for the SFR model and we compare its value with that of the GR case. We also show that there is a relation between the deflection angle of SFR and of Shapiro {\it et al.} \cite{Shapiro} observational result of the deflection angle for very small angles ($\phi\approx 0$) near a fiducial geodesic.
 
 The present work is organized as follows. The structure of the model is given in Section \ref{sec: Basic structure}.  In this framework, the geodesics are  studied and a dynamical analysis is presented in Section \ref{sec: Geodesics}. We also compare our results with GR and discuss the corresponding similarities
 and differences and we give an application to the Newtonian framework.
 A dynamical analysis for the effective potential of this spacetime is provided in the Section \ref{sec: Effective Potential},
 where upon suitable assumptions, the phase portraits
 of both models (SFR and GR) are presented.
In sec. \ref{sec: Deflection}, we study the deflection angle of the SFR spacetime and we compare it with very small values of the deflection angle of Shapiro {\it et al.}~\cite{Shapiro}.
  %, where we derive an interesting relation}. 
  Finally, the conclusions of our study
  and some possible directions for future exploration 
  are presented in Section \ref{sec: Conclusions}.
  
	\section{Basic structure of the model}\label{sec: Basic structure}
	
	In this section, we briefly present the underlying geometry of the SFR gravitational model, as well as the field equations for the SFR metric. The solution of these equations for this metric is presented at the end of the section. An extended study of this model can be found in \cite{Triantafyllopoulos:2020vkx,Triantafyllopoulos:2020ogl}.
	The Lorentz tangent bundle $TM$ is a fibered 8-dimensional manifold with local coordinates $\{x^\mu,y^\alpha\}$ where the indices of the $x$ variables are $\kappa,\lambda,\mu,\nu,\ldots = 0,\ldots,3$ and the indices of the $y$ variables are  $\alpha,\beta,\ldots,\theta = 4,\ldots,7$.
	The tangent space at a point of $TM$ is spanned by the so-called adapted basis
	$\{E_A\} = \,\{\delta_\mu,\dot\partial_\alpha\} $ with
\begin{equation}
\delta_\mu = \dfrac{\delta}{\delta x^\mu}= \pder{}{x^\mu} - N^\alpha_\mu(x,y)\pder{}{y^\alpha} \label{delta x}
\end{equation}
and
\begin{equation}
\dot \partial_\alpha = \pder{}{y^\alpha}
\end{equation}
where $N^\alpha_\mu$ are the components of a nonlinear connection $N=N^{\alpha}_{\mu}(x,y)\,\de x^{\mu}\otimes \pdot{\alpha}$. %\frac{\partial}{\partial y^{\alpha}}$.

The nonlinear connection induces a split of the total space $TTM$ into a horizontal distribution $T_HTM$ and a vertical distribution $T_VTM$. The above-mentioned split is expressed with the Whitney sum:
\begin{equation}
TTM = T_HTM \oplus T_VTM
\end{equation}
The anholonomy coefficients of the nonlinear connection are defined as
\begin{equation}\label{Omega}
\Omega^\alpha_{\nu\kappa} = \dder{N^\alpha_\nu}{x^\kappa} - \dder{N^\alpha_\kappa}{x^\nu}
\end{equation}
	A Sasaki-type metric 
\cite{Miron:1994nvt, Vacaru:2005ht}
	$\Gd$ on $TM$ is:
	\begin{equation}
		\mathcal{G} = g_{\mu\nu}(x,y)\,\mathrm{d}x^\mu \otimes \mathrm{d}x^\nu + v_{\alpha\beta}(x,y)\,\delta y^\alpha \otimes \delta y^\beta \label{bundle metric}
	\end{equation}
	where we have defined the metrics $g_{\mu\nu}$ and $v_{\alpha\beta}$ to be pseudo-Finslerian.
	
	A pseudo-Finslerian metric $ f_{\alpha\beta}(x,y) $ is defined as one that has a Lorentzian signature of $(-,+,+,+)$ and that also obeys the following form:
	\begin{align}
		f_{\alpha\beta}(x,y) = \pm\frac{1}{2}\pdder{F^2}{y^\alpha}{y^\beta} \label{Fg}
	\end{align}
	where the function $F$ satisfies the following conditions  \cite{Miron:1994nvt}:
	\begin{enumerate}
		\item $F$ is continuous on $TM$ and smooth on  $ \widetilde{TM}\equiv TM\setminus \{0\} $, i.e., the tangent bundle minus the null set $ \{(x,y)\in TM | F(x,y)=0\}$ . \label{finsler field of definition}
		\item $ F $ is positively homogeneous of first degree on its second argument:
		\begin{equation}
			F(x^\mu,ky^\alpha) = kF(x^\mu,y^\alpha), \qquad k>0 \label{finsler homogeneity}
		\end{equation}
		\item The form 
		\begin{equation}
			f_{\alpha\beta}(x,y) = \dfrac{1}{2}\pdder{F^2}{y^\alpha}{y^\beta} \label{finsler metric} 
		\end{equation}
		defines a non-degenerate matrix: \label{finsler nondegeneracy}
		\begin{equation}
			\det\left[f_{\alpha\beta}\right] \neq 0 \label{finsler nondegenerate}
		\end{equation}
	\end{enumerate}
	where the plus-minus sign in  \eqref{Fg} is chosen so that the metric has the correct signature.\\[7pt]
	
In the following, we choose a non-linear connection with the following form:
	\begin{equation}\label{Nconnection}
		N^\alpha_\mu = \frac{1}{2}y^\beta g^{\alpha\gamma}\partial_\mu g_{\beta\gamma}
	\end{equation}
	The metric tensor $v_{\alpha\beta}$  is derived from \eqref{RandersL} by using \eqref{Fg}, after omitting higher order terms $O(A^2)$:
	\begin{equation}\label{vab}
		v_{\alpha\beta}(x,y) = g_{\alpha\beta}(x) + h_{\alpha\beta}(x,y)   
	\end{equation}
	where 
	\begin{align}\label{hab}
		h_{\alpha\beta} = \frac{1}{\tilde{a}}(A_{\beta}g_{\alpha\gamma}y^\gamma + A_{\gamma}g_{\alpha\beta}y^\gamma + A_{\alpha}g_{\beta\gamma}y^\gamma) + \frac{1}{\tilde{a}^3}A_{\gamma}g_{\alpha\epsilon}g_{\beta\delta}y^\gamma y^\delta y^\epsilon
	\end{align}
	with $\tilde{a} = \sqrt{-g_{\alpha\beta}y^{\alpha}y^{\beta}}$.
	The total metric defined in the previous steps is called the \textit{Schwarzschild-Finsler-Randers} (SFR) metric. As we can see, the term $h_{\alpha\beta}(x,y)$ can be considered as a perturbation of the Schwarzschild metric since $|A_\gamma(x)|\ll 1$.

	The nonzero coefficients of a canonical and distinguished $d-$connection $\mathcal D$ on $TM$ are:
	%can be found in \cite{miron-watanabe-ikeda 1987}:
	\begin{align}
	L^\mu_{\nu\kappa} & = \frac{1}{2}g^{\mu\rho}\left(\delta_kg_{\rho\nu} + \delta_\nu g_{\rho\kappa} - \delta_\rho g_{\nu\kappa}\right) \label{metric d-connection 1}  \\
	L^\alpha_{\beta\kappa} & = \dot{\partial}_\beta N^\alpha_\kappa + \frac{1}{2}v^{\alpha\gamma}\left(\delta_\kappa v_{\beta\gamma} - v_{\delta\gamma}\,\dot{\partial}_\beta N^\delta_\kappa - v_{\beta\delta}\,\dot{\partial}_\gamma N^\delta_\kappa\right) \label{metric d-connection 2}  \\
	C^\mu_{\nu\gamma} & = \frac{1}{2}g^{\mu\rho}\dot{\partial}_\gamma g_{\rho\nu} \label{metric d-connection 3} \\
	C^\alpha_{\beta\gamma} & = \frac{1}{2}v^{\alpha\delta}\left(\dot{\partial}_\gamma v_{\delta\beta} + \dot{\partial}_\beta v_{\delta\gamma} - \dot{\partial}_\delta v_{\beta\gamma}\right) \label{metric d-connection 4}
	\end{align}
	See Appendix \ref{sec:d-connection} for more details.
	
	The field equations for our model have been derived in previous works and can be found in Appendix \ref{sec:field_eqs}. The solution of the field equations \eqref{feq1}, \eqref{feq2} and \eqref{feq3} to first order in $A_\gamma(x)$ in vacuum ($T_{\mu\nu} = Y_{\alpha\beta} = \mathcal Z^\kappa_\alpha = 0$) is  \cite{Triantafyllopoulos:2020vkx}:
	\begin{equation}\label{Asolution}
		A_\gamma(x) = \left[\tilde A_0 \left(1-\frac{R_S}{r}\right) ^{1/2}, 0, 0, 0 \right]=\left[\tilde A_0 f^{1/2}, 0, 0, 0 \right]
	\end{equation}
	with $\tilde A_0$ a constant.
	While this is an approximate solution, it will be sufficient
	for our purposes given the assumption $|A_\gamma(x)|\ll 1$.
	
\section{Geodesics}\label{sec: Geodesics}
\label{geodesics}
%	A Sasaki-type metric $\Gd$ on $TM$ is given by:
%	\begin{equation}
%		\mathcal{G} = g_{\mu\nu}(x,y)\,\mathrm{d}x^\mu \otimes \mathrm{d}x^\nu + v_{\alpha\beta}(x,y)\,\delta y^\alpha \otimes \delta y^\beta \label{bundle metric}
	%\end{equation}
	{In this section, we will study the geodesics of the SFR and perform a dynamical analysis. We compare our results with the corresponding ones of GR.}
	%\\[7pt]
From the definition of the metric function \eqref{RandersL} we have:
\begin{equation}
\label{metric function}
F(x,dx) = \left(-g_{\mu\nu}(x)dx^{\mu}dx^{\nu}\right)^{1/2} + A_{\gamma}(x)dx^{\gamma}
\end{equation}
where $g_{\mu\nu}(x)$ is the Schwarzschild metric and $A_{\gamma}(x)$ is a one-form vector field with $|A_\gamma(x)|\ll 1$.\\[7pt]
By using the rel.~\eqref{Schwarzchild} the rel.~\eqref{metric function} is written as:
\begin{equation}
    	F(x,dx) = \Big[fdt^2 - \frac{dr^2}{f} - r^2 d\theta^2 - r^2 \sin^{2}\theta\, d\phi^2\Big]^{1/2} + A_{\gamma}(x)dx^{\gamma}
    	\label{metrical function}
\end{equation}

We define the Lagrangian 
\begin{equation}\label{Lagrangian}
L(x,\dot x)=F(x,\dot x)=\Big[f\dot{t^{2}} - \frac{\dot{r^{2}}}{f}-r^{2}\dot\theta^{2}-r^{2}sin^{2}\theta\dot{\phi^2}\Big]^{1/2}+\tilde{A_{0}}f^{1/2}\dot{t}
\end{equation}
where we denote $\dot{x}=\frac{dx}{d\tau}$ and we have used Eqs.~\eqref{metrical function} and~\eqref{Asolution}.
%\\[7pt]
From the Euler-Lagrange equations
\begin{equation}
\frac{d}{d\tau}\frac{\partial L}{\partial \dot{x}^{\mu}}=\frac{\partial L}{\partial x^{\mu}}    
\end{equation}
we find the equations for the geodesics:
\begin{equation}
\ddot{x}^{\lambda}+\Gamma^{\lambda}_{\mu\nu}\dot{x}^{\mu}\dot{x}^{\nu}+g^{\kappa\lambda}\Phi_{\kappa\mu}\dot{x}^{\mu}=0  
\end{equation}
where $\Gamma^{\lambda}_{\mu\nu}$ are the Christoffel symbols of Riemann geometry, $\dot{x}^{\mu}=\frac{dx^{\mu}}{d\tau}$ and $\Phi_{\kappa\mu}=\partial_{\kappa}A_{\mu}-\partial_{\mu}A_{\kappa}$ and $A_{\mu}$ is the solution rel.\eqref{Asolution}. We notice that from the definition of $\Phi_{\kappa\mu}$ we get a rotation form of geodesics. If $A_{\mu}$ is a gradient of a scalar field, $A_{\mu}=\pder{\Phi}{x^{\mu}}$ then $\Phi_{\kappa\mu}=0$ and the geodesics of our model are identified with the Riemannian ones.
\\[7pt]
The geodesics of our model can then be explicitly written in the form:
\begin{align}
 	&\ddot t + \frac{1-f}{rf}\dot r \dot t =-\tilde{A}_{0}\dot{r}\frac{f^{-3/2}(1-f)}{2r}
\label{geodesics0}\\
&\ddot r + \frac{f(1-f)}{2r} \dot t^2 - \frac{1-f}{2rf} \dot r^2 - rf \big( \dot \theta^2 + \sin^2 \theta \dot  \phi^2 \big)=-\tilde{A}_{0}\dot{t}\frac{f^{1/2}(1-f)}{2r}
\label{geodesics1}\\
&\ddot \theta + \frac{2}{r} \dot \theta \dot r - \frac{1}{2}\sin 2\theta \, \dot \phi^2=0
\label{geodesics2}\\
&\ddot \phi + \frac{2}{r} \dot \phi \dot r + 2\cot \theta \, \dot \theta \dot \phi=0
\label{geodesics3}
\end{align}
From the relations \eqref{geodesics0}-\eqref{geodesics3}, we notice that the first two dynamical equations involve a contribution of extra terms 
particular to the SFR spacetime while the last two relations are the same as in GR. From the above mentioned relations, we notice that the Riemannian  geodesics are affected by a contribution that involves the curl of the force field in the SFR spacetime which can be interpreted as extra energy for the content of GR. We can notice that the geodesics of SFR are destroyed at the singular value $r=0$ as in the GR case, this can be seen from the relations \eqref{geodesics0}-\eqref{geodesics3}. In our case, the singularity is inherited from the Schwarzschild spacetime of GR. Moreover, it is evident
from their dynamical form that the equations are
not meaningful for $r \leq R_S$ in the context of
the presently considered Schwarzschild metric, hence
we only consider radial displacements past this singular
point.

We now make a key assumption regarding the angular 
dependence of the model. Namely,
by using $\theta=\frac{\pi}{2}$ we notice that Eq.~\eqref{geodesics2} is satisfied and equations \eqref{geodesics0}, \eqref{geodesics1} and \eqref{geodesics3} can be written as:
\begin{align}
 	&\ddot t + \frac{1-f}{rf}\dot r \dot t =-\tilde{A}_{0}\dot{r}\frac{f^{-3/2}(1-f)}{2r}
\label{geodesics0v2}\\
&\ddot r + \frac{f(1-f)}{2r} \dot t^2 - \frac{1-f}{2rf} \dot r^2 - rf\dot\phi^2 =-\tilde{A}_{0}\dot{t}\frac{f^{1/2}(1-f)}{2r}
\label{geodesics1v2}\\
&\ddot \phi + \frac{2}{r} \dot \phi \dot r=0
\label{geodesics3v2}
\end{align}
From Eq.\eqref{geodesics3v2} we find:
\begin{equation}
r^{2}\dot{\phi}=J=const.
\label{angular momentum}
\end{equation}
where $J$ is the angular momentum and the relevant equation 
represents its conservation law.
  If we use the relation $f'=\frac{1-f}{r}$ where $f=1-\frac{2GM}{r}$ and the Leibniz chain-rule $\frac{d}{d\tau}=\frac{dr}{d\tau}\frac{d}{dr}=\dot r \frac{d}{dr}$, then Eq.~\eqref{geodesics0v2} can be written as:
  \begin{equation}
      f\ddot t + \frac{df}{d\tau}\dot t  = -\tilde{A}_{0}\frac{df^{1/2}}{d\tau}
  \end{equation}
which, in turn, gives us
\begin{equation}\label{energy0}
f\dot{t}+\tilde{A}_{0}f^{1/2}  = \mathcal{E}_R=const.  
\end{equation}
where $\mathcal{E}_R$ is the energy of the particle moving along the geodesic.
We notice 
%from eq.\eqref{energy}
that the first term constitutes the energy for a particle moving along the geodesics in general relativity, $\mathcal{E}_{GR}=f\dot{t}$ and we can rewrite the %rel.\eqref{energy} as:
relevant expression as:
\begin{equation}\label{energy-Randers}
\mathcal{E}_{GR}+\tilde{A}_{0}f^{1/2}=\mathcal{E}_{R}.    
\end{equation}

By using Eq.~\eqref{geodesics1v2} with \eqref{angular momentum}, 
and \eqref{energy0}, 
we arrive at the (effectively one-degree-of-freedom) radial equation:
\begin{equation}
 \ddot r + \frac{1-f}{2rf}(\mathcal{E}_{R}^{2}-\dot{r}^{2}) -\frac{fJ^{2}}{r^{3}}=\tilde{A}_{0}\mathcal{E}_{R}\frac{f^{-1/2}(1-f)}{2r}  
 \label{radial1}
\end{equation}
where we omitted $O(\tilde A_{0}^{2})$ terms.
As before, we use the relation $f'=\frac{1-f}{r}$ in \eqref{radial1} to
bring it to the equivalent form:
\begin{equation}
\label{radial2}
\ddot r + \frac{f'}{2f}(\mathcal{E}_{R}^{2}-\dot r^{2})-\frac{fJ^{2}}{r^{3}}=\frac{\tilde A_{0}\mathcal{E}_{R}}{2}f^{-1/2}f' 
\end{equation}

We can further simplify the Eq.~\eqref{radial2} by using the Leibniz chain-rule $\frac{d}{d\tau}=\frac{dr}{d\tau}\frac{d}{dr}=\dot r \frac{d}{dr}$ and upon deriving the first integral of the motion, we obtain:
\begin{equation}
\dot{r}^{2} + f\left(\frac{J^{2}}{r^{2}}+\epsilon\right)+2\tilde{A}_{0}\mathcal{E}_{R}f^{1/2}=\mathcal{E}_{R}^{2}
\label{radial3}
\end{equation}
where $\epsilon$ is a constant and for $\epsilon = 0$ we have null geodesics.
It is important to indicate here that for each
value of $\epsilon$, we obtain a different curve of
this ``first integral'' of Eq.~(\ref{radial3}) and the
inclusion of all of the admissible ($r$,$\dot{r}$) curves
will provide us with the phase portraits presented below.
The first two terms from \eqref{radial3} constitute the total energy in general relativity (GR), $\mathcal{E}^{2}_{GR}=\dot{r}^{2} + f(\frac{J^{2}}{r^{2}}+\epsilon)$ and the third term emerges from the structure of SFR spacetime and its energetic contribution.  
Therefore Eq.~\eqref{radial3} can be written as:
\begin{equation}
\label{energy}
\mathcal{E}^{2}_{GR}+2\tilde{A}_{0}\mathcal{E}_{R}f^{1/2}=\mathcal{E}_{R}^{2}    
\end{equation}
Eq.~\eqref{energy} shows that the term $A_{\gamma}(x)$ from \eqref{Asolution} provides an additional energy contribution to the system of GR.

For a particle moving along the geodesics in the SFR spacetime with metric function $F_v(x,dx)$ (rel. \ref{metrical function}),
we have extra gravitational effect (energy) compared to GR. An amount of energy comes from the gravitational field of total space in the SFR model. Therefore, from the rel. \eqref{energy0} and \eqref{energy}, the second term $\tilde{A}_{0}f^{1/2}$ can be interpreted as the difference of energy between $\mathcal{E}_{R}$ and $\mathcal{E}_{GR}=f\dot{t}$. \\[4pt]
%Equivalently, this term is also included in the rel.~(42).

\noindent
\textit{\textbf{Remark 1}}

The action of a vector field in the pseudo-Riemannian structure of space-time can give a vector-dependent gravitational field of Finsler-Randers (FR) type which can describe the asymmetry and the locally anisotropic form of spacetime that the Riemannian geometry is unable to provide. In this approach, because of broken Lorentz symmetry, particles and forces interact with this vector field.
%In addition, FR models can be related to the chiral fields in cosmology for different eras of the Universe.}\\[4pt]

In the following, we give an application for the weak gravitational field where we show that the term $A_{0}(r)$ in the Eq.~\eqref{Asolution} is connected to the gravitational potential $\Phi$.\\[7pt] 

\noindent
{\bf \textit{Application to the Newtonian limit.}}\\[7pt]
In the Newtonian gravitational theory, we can consider the second term of rel.\eqref{lagrangian2} of the manuscript in the following form:
\begin{equation}
    f_a dx^a = \frac{GmM}{r^2}dr
\end{equation}
The gravitational potential is given by
\begin{equation}\label{gravitational potential}
    U = \frac{ W}{m} = \frac{1}{m}\int f_a dx^a = \frac{1}{m}\int^r_\infty \frac{GmM}{r^2}dr = -\frac{GM}{r} = \Phi
\end{equation}
where $W$ represents the gravitational potential energy to be done to bring a unit mass m from infinity to a point and $r$ is the distance from a mass $M$. This means that the work is converted to gravitational potential energy.
In addition, from the relations \eqref{Schwarzchild} and \eqref{Asolution}
we have
\begin{equation}\label{g00}
    g_{00} = - \left(1 - \frac{R_S}{r} \right) = -f
\end{equation}
and
\begin{equation}\label{A0(r)}
    A_0(r) = \tilde{A}_0 \left(1 - \frac{R_S}{r} \right)^{1/2} = \tilde A_0 f^{1/2}
\end{equation}
From the  relations \eqref{gravitational potential}, \eqref{g00} and \eqref{A0(r)}, we obtain
\begin{equation}\label{A4_potential}
   A_0(r) = \tilde A_0 \sqrt{-g_{00}} = \tilde A_0 \left(1 - \frac{\Phi}{2} \right)^{1/2} 
\end{equation}
where we have used $R_S = 2GM$. The rel. \eqref{A4_potential} gives a physical interpretation in the geometrical term $A_0(r)$ because of its dependence on the gravitational potential $\Phi$. The term $A_0(r)$ is included in the relations \eqref{energy0}, \eqref{energy-Randers}, \eqref{energy} giving a physical meaning to these equations.\\[4pt]

\noindent
\textbf{\textit{Remark 2}}

The relation \eqref{A4_potential} is also related to gravitational redshift of photons in the SFR spacetime (ref. [2], paragraph 6). In that case we have proved that :
\begin{equation}
\frac{E_{rec}}{E_{emit}}=
\frac{\nu_{rec}}{\nu_{emit}}\approx \tilde{A_{0}}f^{1/2} 
\end{equation}
where $\nu_{rec},\nu_{emit}$ denote the frequencies of receiver and emitter.\\[4pt]

\begin{figure}[H]
\begin{subfigure}{0.5\textwidth}
\includegraphics[scale=0.4]{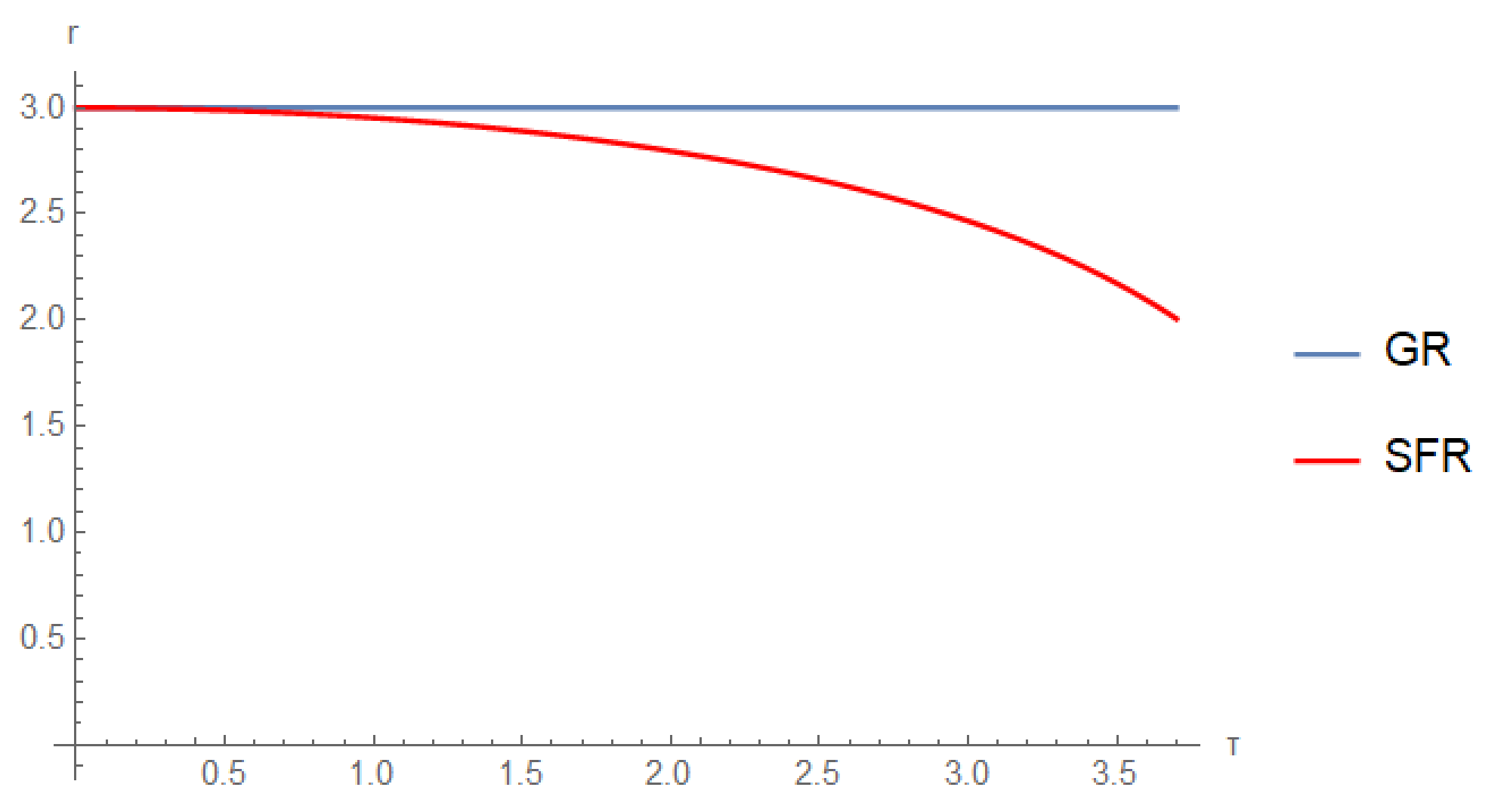} 
\caption{This is an $(\tau,r)$ graph for the geodesics of photons for angular momentum $J=4$ and initial radial distance $r_{0}=3$. The red line shows the SFR geodesics and the blue line the GR geodesics}
\label{graph-sfr-r-s}
\end{subfigure}
\begin{subfigure}{0.45\textwidth}
\includegraphics[scale=0.4]{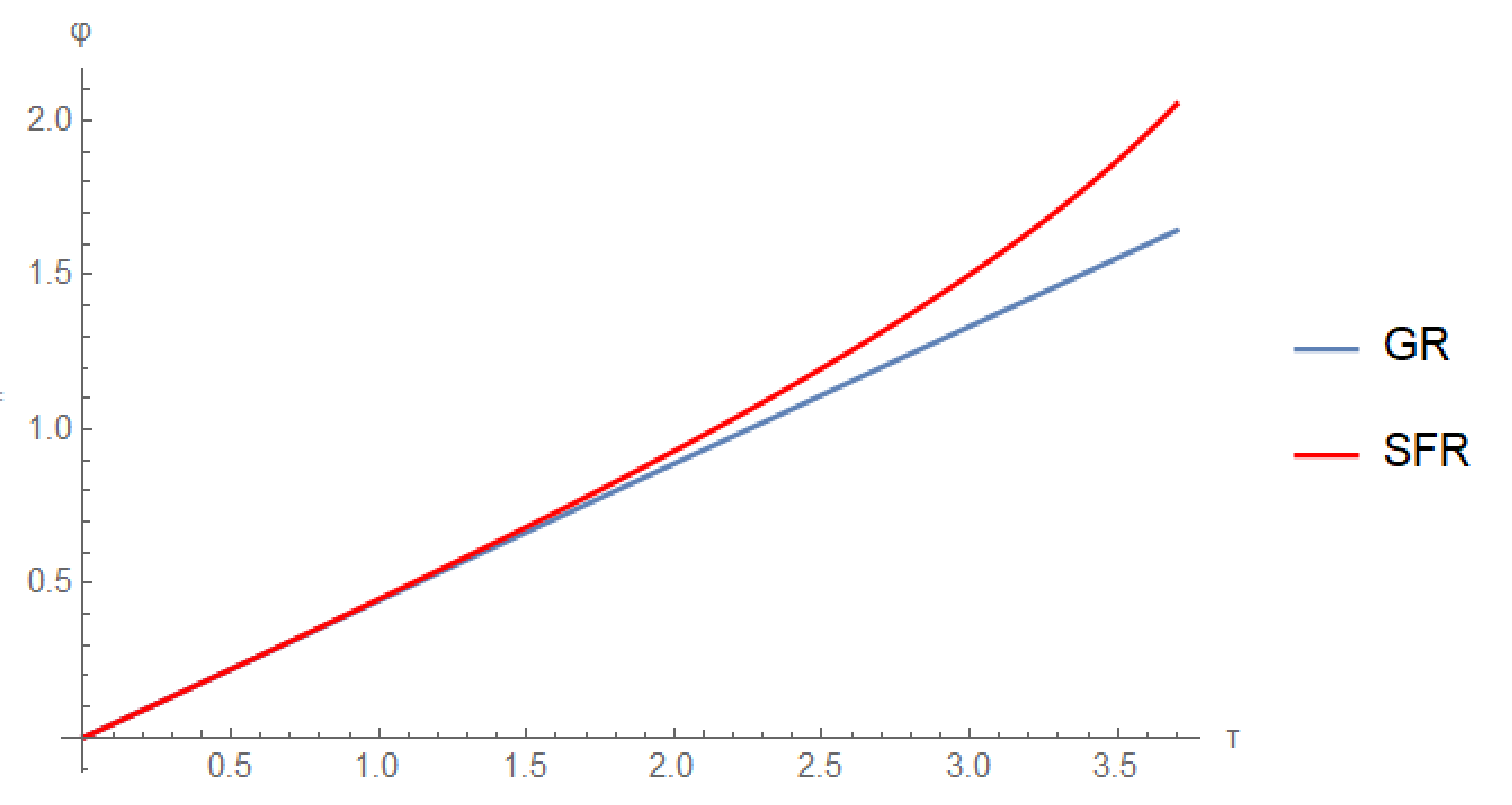}
\caption{This is an $(\tau,\phi)$ graph for the geodesics of photons for angular momentum $J=4$ and initial radial distance $r_{0}=3$. The red line shows the SFR geodesics and the blue line the GR geodesics}
\label{graph-sfr-φ-s}
\end{subfigure}
\caption{These are the graphs for the GR and SFR geodesics of photons
for angular momentum $J = 4$ and initial radial distance $r_0 = 3$}\label{graph}
\end{figure}

\begin{figure}
\includegraphics[scale=0.5]{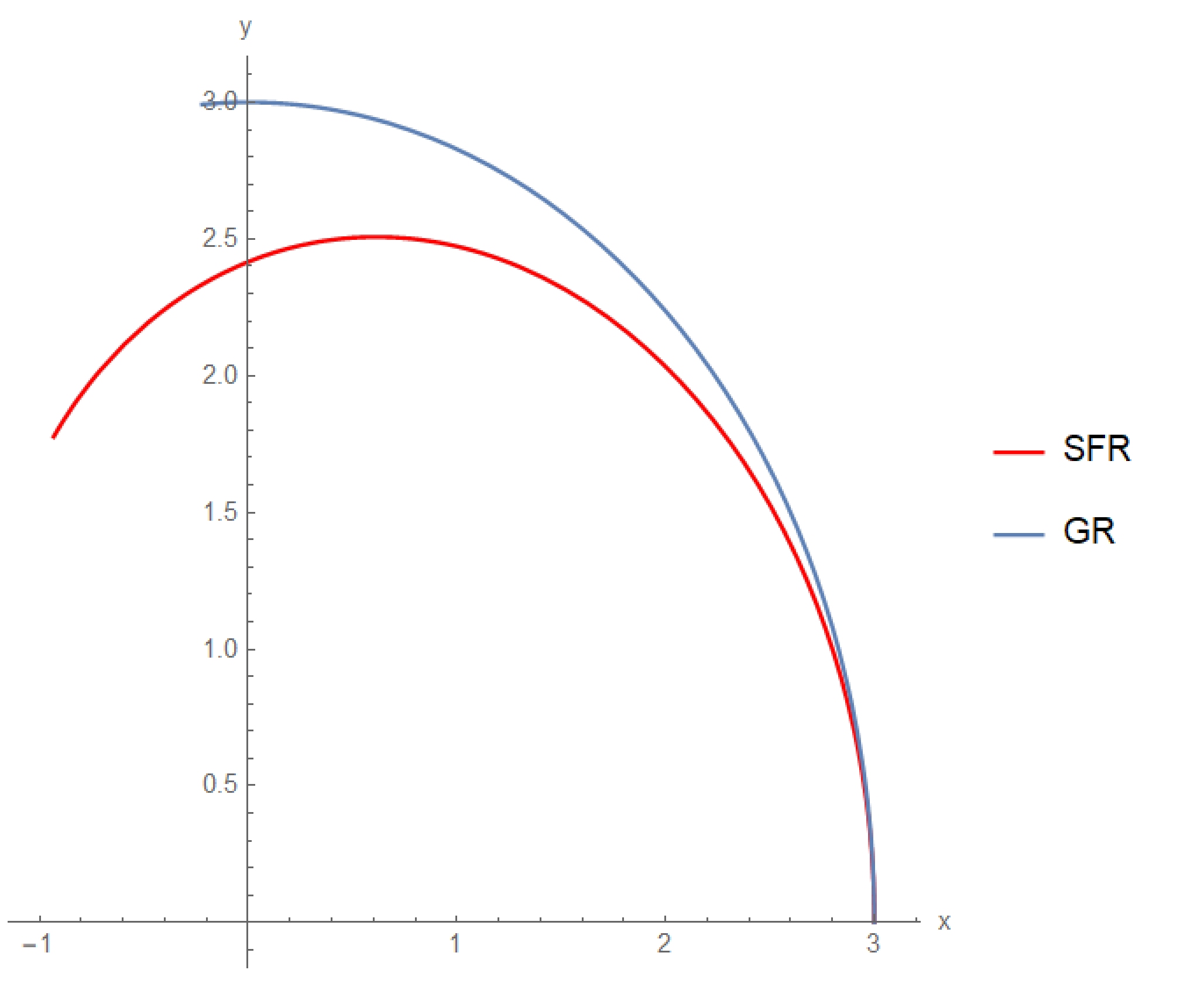}
\caption{This is an $x-y$ graph for the geodesics of photons for angular momentum $J=4$ and initial radial distance $r_{0}=3$. The red line shows the SFR geodesics and the blue line the GR geodesics}
\label{graph-sfr-x-y}
\end{figure}

Below, we give the Figures \ref{graph-sfr-r-s}, \ref{graph-sfr-φ-s} and Fig. \ref{graph-sfr-x-y} for the geodesics of GR and SFR we have obtained by solving the equations \eqref{geodesics0}-\eqref{geodesics3}.
The relevant ordinary differential equations are solved via
a standard solver within Mathematica and $(r,\phi)$ are presented
as a function of $\tau$, while Fig.~\ref{graph-sfr-x-y} presents
the evolution in the original $(x,y)$ plane. In our case, we assume $R_{s}=2$ and initial radial distance $r_{0}=3$, so the photons are found on the photonsphere with $r_{ph}=\frac{3}{2}R_{s}=3$ in the GR case. The deviation between
the trajectories of the SFR and those of the GR is clearly discernible
in both figures.
From Fig.~\ref{graph-sfr-r-s}, we can see that the radial component in the SFR model takes lower values compared to the GR one which remains constant. This difference between the r-components of SFR and GR can be interpreted as the 
increase of the radius of the photonsphere due to the one-form $A_{\gamma}$ as we have shown in \cite{Kapsabelis:2021dpb}. %This leads to the decrease of the radial component photon's radial component 
%in SFR
%because the orbit radially decreases in the SFR model 
%while for $r_{0}=3$ the orbit preserves its radial distance in GR.}
This leads the orbit of the photon to fall inside the event horizon because the initial distance $r_{0}=3$ and energy are not sufficient to allow circular orbits of the photonsphere.
That means for an
%stable 
% It is not really stable!
orbit with $r$ constant in the SFR model, the particle needs more energy compared to the GR case. %in order to reach the larger radial distance that is necessary .} Similarly, from Fig.~\ref{graph-sfr-φ-s} the angle $\phi$ takes larger values because the orbit is not $\tau$-independent
%stable
% Again it is not a matter of stability
%and the particles/photons fall inside the event horizon. 
In Fig.~\ref{graph-sfr-x-y}, the geodesics of GR and SFR are depicted. In the case of GR, the photons move in circular orbits around the black hole. In the SFR model, the photons follow a spiral orbit and fall inside the event horizon. 

It is important to remind the reader here that underlying these results
is the key assumption of $\theta=\frac{\pi}{2}$ which allows the 
reduction of the model to an \textit{effective single degree-of-freedom
system}. It is important in future work to consider how deviations
from this equilibrium value (and the corresponding incorporation
of the full dynamical system) may affect the conclusions presented
above. However, as the latter is outside the scope of the present
study, we now focus on the further analysis of the effective 
potential of the SFR model and its implications for the
phase portrait of the relevant system.

\section{Effective Potential of SFR model}\label{sec: Effective Potential}

\begin{figure}[H]
\begin{subfigure}{0.5\textwidth}
    \includegraphics[scale=0.4]{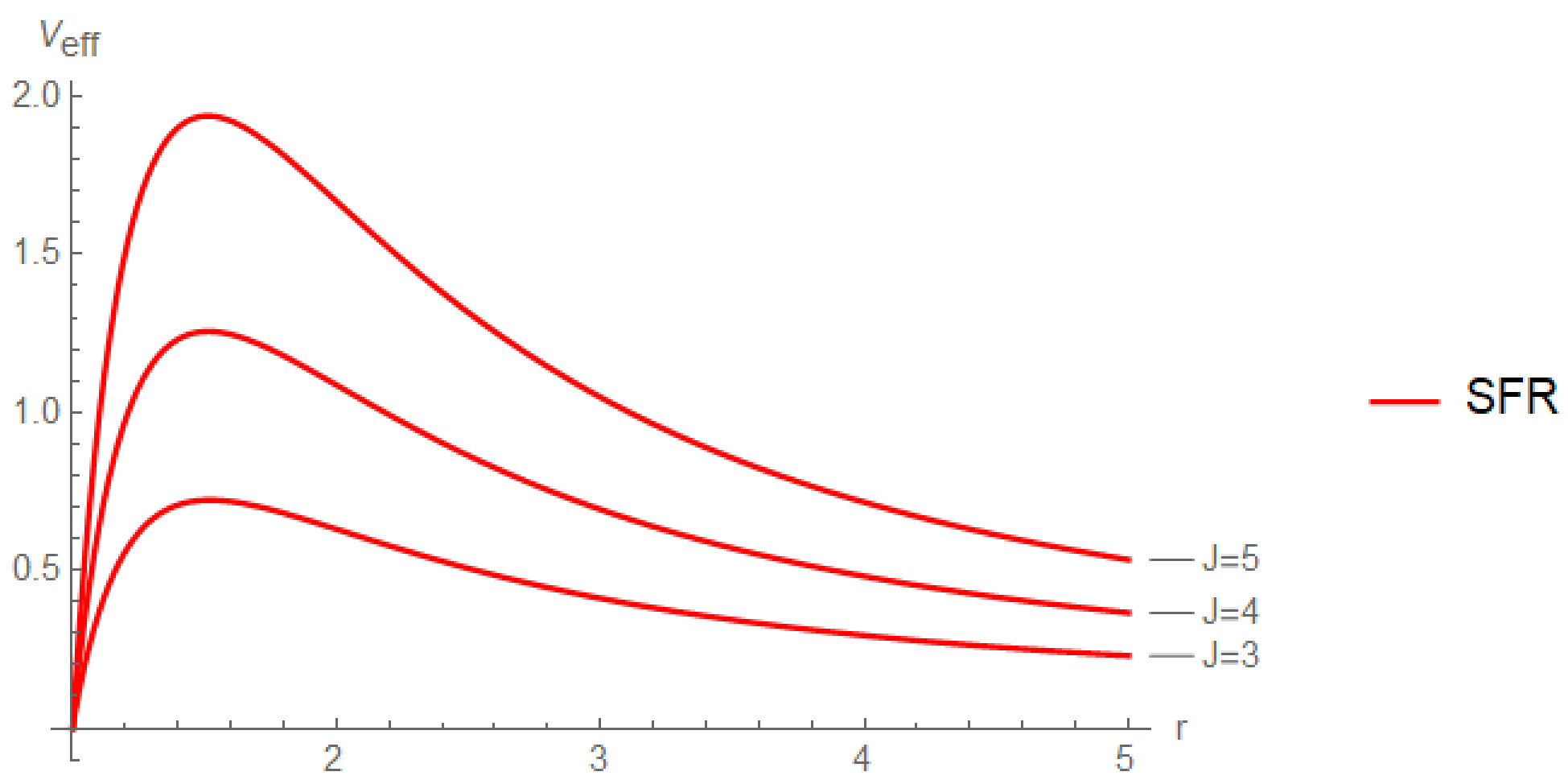}
    \caption{Graph for the $V_{eff}(r)$ in the SFR model for\\ angular momentum $J=3$, $J=4$ and $J=5$}
    \label{VeffSFR}
\end{subfigure}
\begin{subfigure}{0.5\textwidth}
    \includegraphics[scale=0.4]{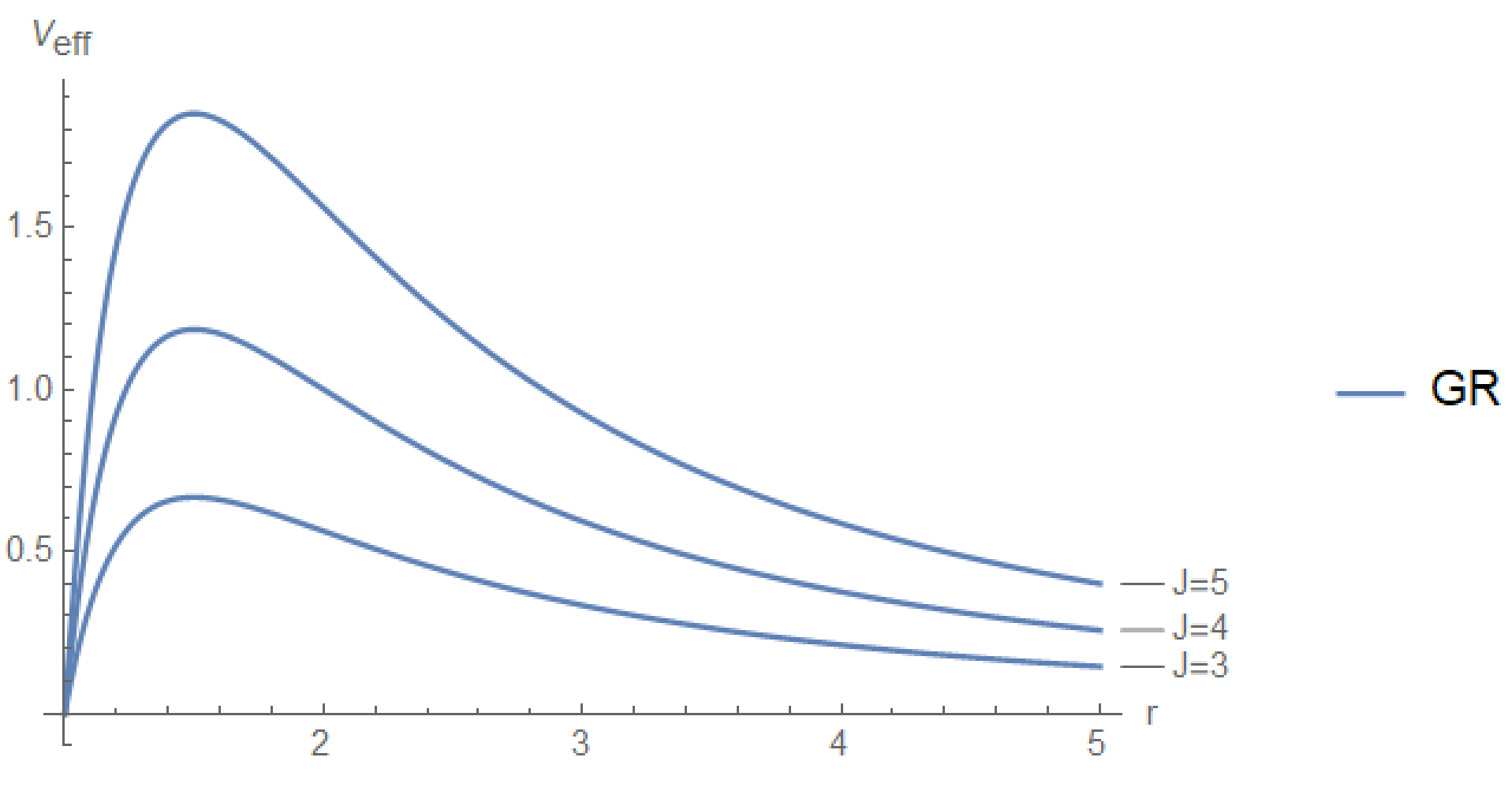}
    \caption{Graph for the $V_{eff}(r)$ in the GR model for\\ angular momentum $J=3$, $J=4$ and $J=5$}
    \label{VeffGR}
\end{subfigure}
\begin{subfigure}{0.5\textwidth}
\includegraphics[scale=0.4]{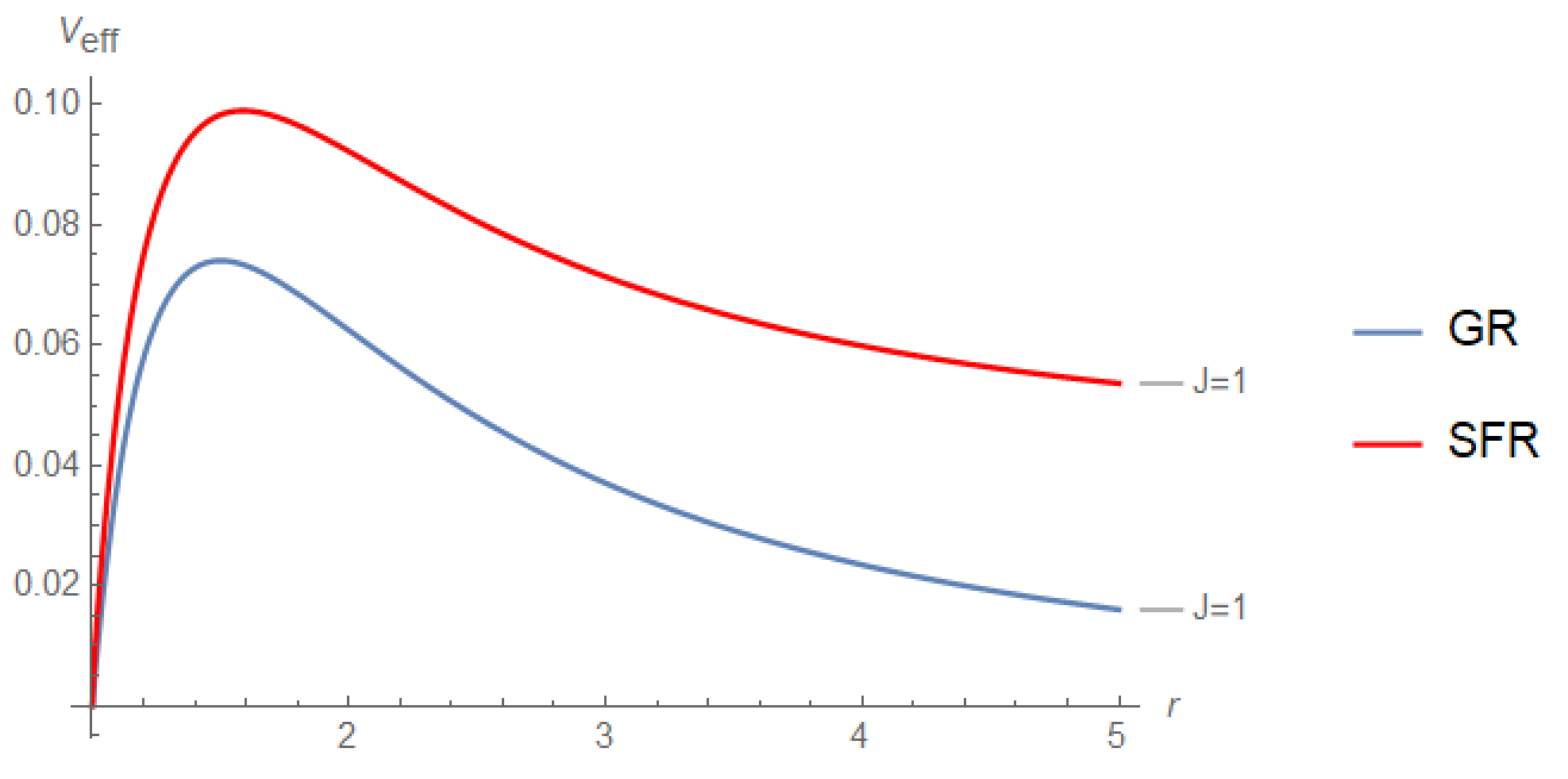} 
\caption{Graph for the $V_{eff}(r)$ in GR (blue line)\\ and SFR (red line) for angular momentum $J=1$}
\label{VeffSFRGR1}
\end{subfigure}
\begin{subfigure}{0.5\textwidth}
\includegraphics[scale=0.4]{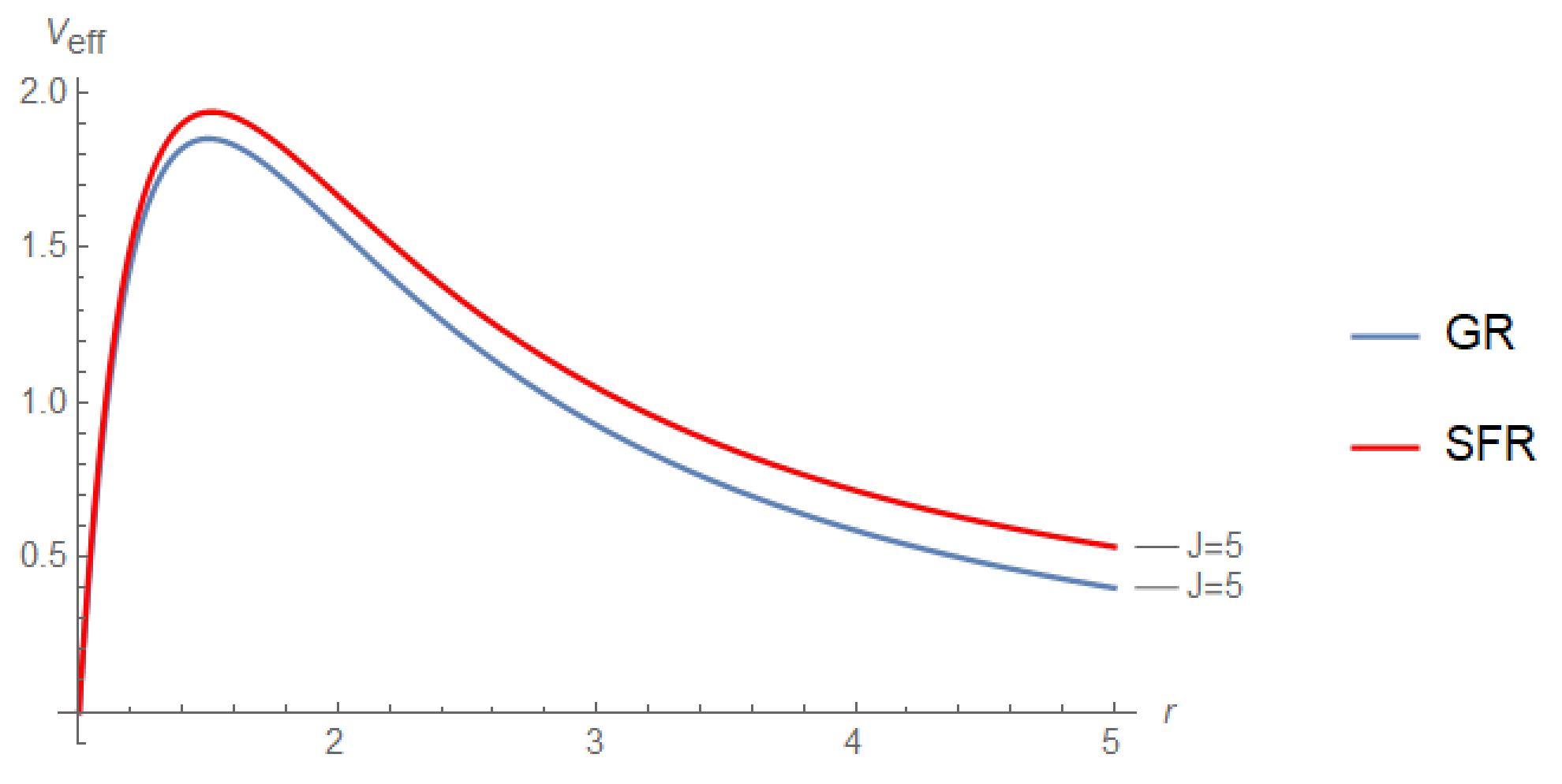}
\caption{This is a graph for the $V_{eff}(r)$ in GR (blue line) and SFR (red line) for angular momentum $J=5$}
\label{VeffSFRGR5}
\end{subfigure}
\caption{These are the graphs for the $V_{eff} (r)$ in the GR and SFR models
for various angular momenta}\label{fig:phase}
\end{figure}

In this section, we will study the effective potential of the SFR model and compare it with the effective potential of GR.
%\\[7pt]
The equation of the energy in GR reads:
\begin{equation}
\label{energy eq GR}
 \dot{r}^{2} + f\left(\frac{J^{2}}{r^{2}}+\epsilon\right)=\mathcal{E}_{GR}^{2} \end{equation}
We see from \eqref{energy eq GR} that the effective potential 
energy landscape is given by:
\begin{equation}
    V_{eff,GR}=\frac{1}{2}f \left(\frac{J^{2}}{r^{2}}+\epsilon \right)
\end{equation}\\
In Fig.\ref{VeffGR}, we show the graph for the effective potential in GR for angular momentum $J=3,J=4$ and $J=5$ to examine its variation for
different values of the angular momentum.
Our effective potential (here and in what follows 
in Fig.~\ref{VeffSFR}) solely
bears a maximum at a finite $r>0$. Based on the general
theory of dynamical systems for the conservative
type of problems considered herein, the relevant 
fixed point will be a saddle, as will be confirmed in
the corresponding phase portraits below (where the
attracting direction of the stable manifold and the
repelling direction of the unstable manifold will be
evident).
%\\[7pt]

We now recall the key difference (and associated additional contribution)
to the energetics of the SFR model. In particular,
the energy equation for the latter, derived from Eq.~\eqref{radial2}, is given as: 
\begin{equation}
\dot{r}^{2} + f\left(\frac{J^{2}}{r^{2}}+\epsilon \right) + 2\tilde{A}_{0}\mathcal{E}_{R}f^{1/2}= \mathcal{E}_{R}^{2}
\label{energy eq SFR}
\end{equation}
In \eqref{energy eq SFR} the effective potential can be written in the form

\begin{equation}
\label{VeffSFReq}
  V_{eff,SFR}=\frac{1}{2}f \left(\frac{J^{2}}{r^{2}}+\epsilon \right)+ \tilde{A}_{0}\mathcal{E}_{R}f^{1/2}
\end{equation}\\
The graph for the effective potential in the SFR model $(V_{eff},r)$ is depicted in Fig.~\ref{VeffSFR}, in this case
for different values of angular momentum.

In Fig. \ref{VeffSFRGR1} and \ref{VeffSFRGR5} we show the effective potentials of the SFR and GR models comparing the two for $J=1$ and $J=5$.
%In the Fig. \ref{VeffSFRGR1} and \ref{VeffSFRGR5} we present the comparison of the effective potentials for GR and SFR. 
As we can see in Fig.~\ref{VeffSFRGR1}, the difference between GR and SFR is bigger than that of Fig.~\ref{VeffSFRGR5}. Notably, when the contribution
of the angular momentum is weaker, the difference between the two models
is more substantial/clearly discernible. When the angular momentum becomes
large, the relevant difference is rather weak and the $V_{eff}$
of the two models become proximal.

%It is ought to the second term of rel.\eqref{VeffSFReq} where an amount of energy is added to the total effective potential.

In Fig.~\ref{fig:phase} %{-sfr}-\ref{phase-gr} and also~\ref{phase-sfr-gr},
we observe the phase portraits associated with the
effective potentials depicted above. These phase portraits
reflect the existence of an energy barrier whose precise
height depends on the value of the angular momentum.
Energies below this barrier height result in reflection from
the outside and trapping from the inside. On the other hand,
energies higher than those of the barrier result in
reaching the Schwarzschild radius (if the particle is
coming from the outside) or reaching infinity (if the
particle is moving outward from the inside). The
latter figure demonstrates the differences between the two
phase portraits which are quantitative but not qualitative.
%These differences are more pronounced the smaller the value of
%the angular momentum $J$.

%\subsection{DYNAMICAL ANALYSIS OF THE GEODESICS}
\begin{figure}[H]
\begin{subfigure}{0.55\textwidth}
\includegraphics[scale=0.5]{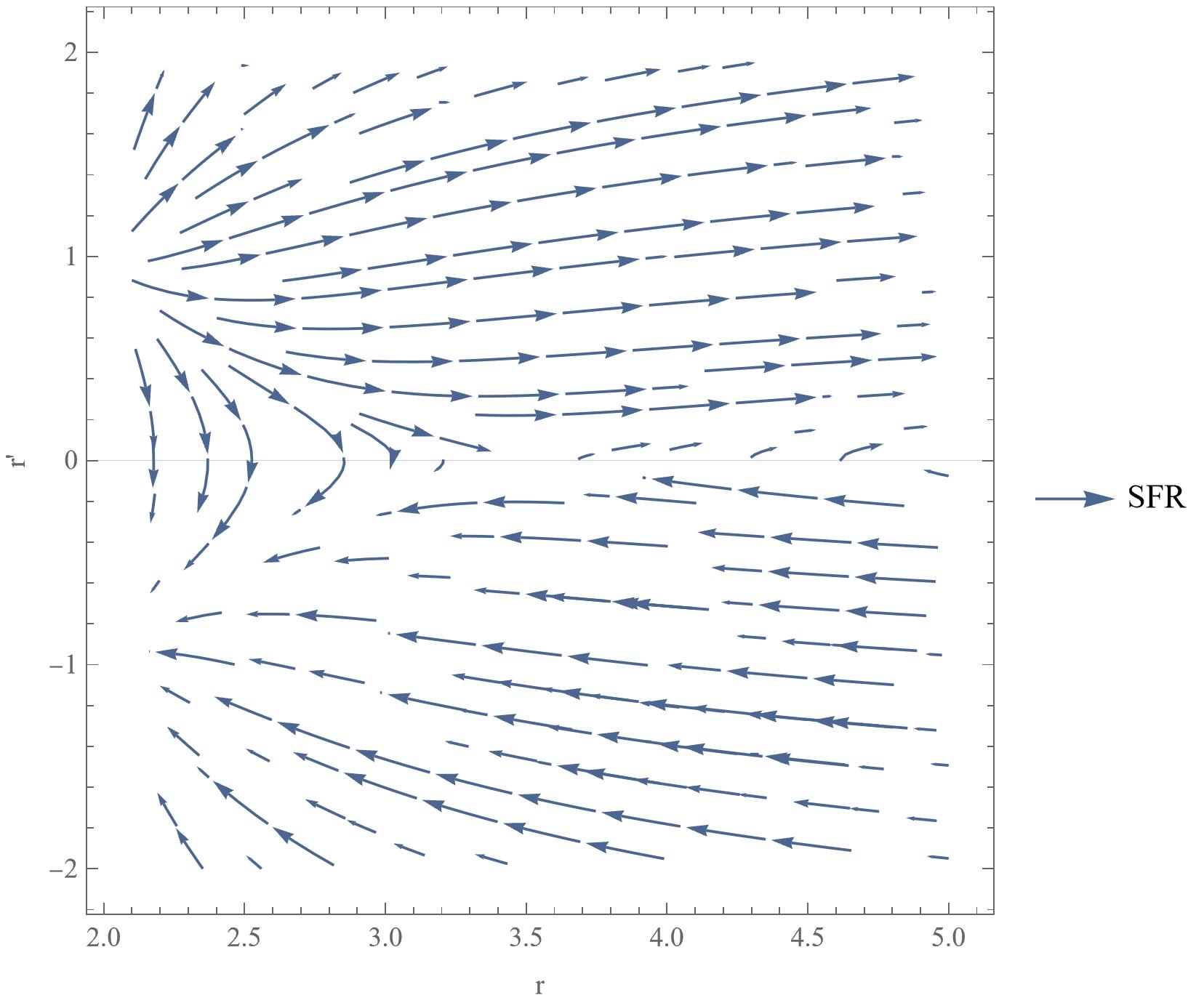} 
\caption{This is a phase plot for the radial geodesics of \\SFR,
representing the trajectories in $(r, \dot{r})$ space}
\label{phase-sfr}
\end{subfigure}
\begin{subfigure}{0.55\textwidth}
\includegraphics[scale=0.5]{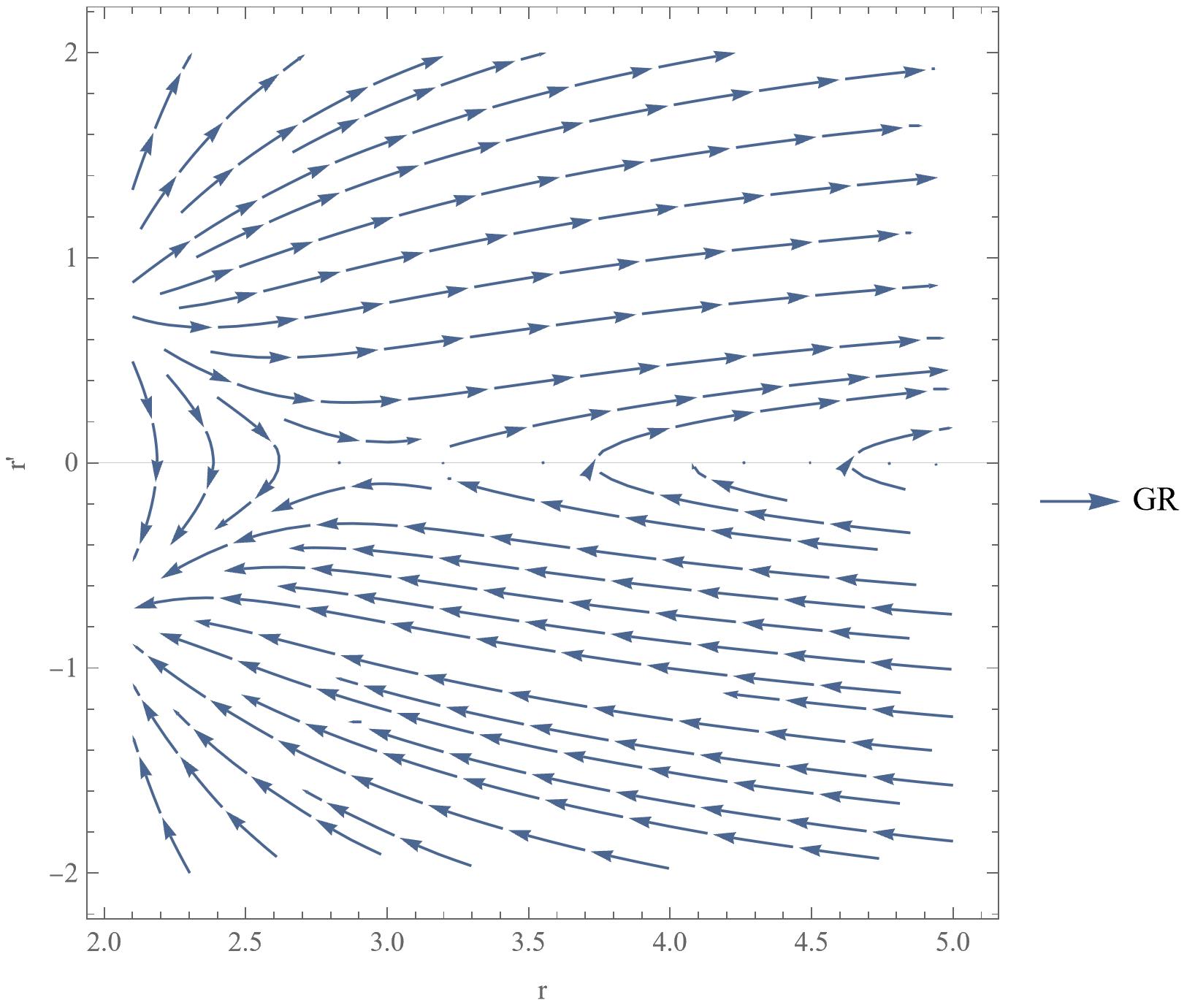} 
\caption{This is a phase plot for the radial geodesics of GR
representing the trajectories in $(r, \dot{r})$ space}
\label{phase-gr}
\end{subfigure}
\caption{These are phase plots for the radial geodesics of the GR and
SFR models}
\end{figure}

\begin{figure}[h]
\includegraphics[scale=0.5]{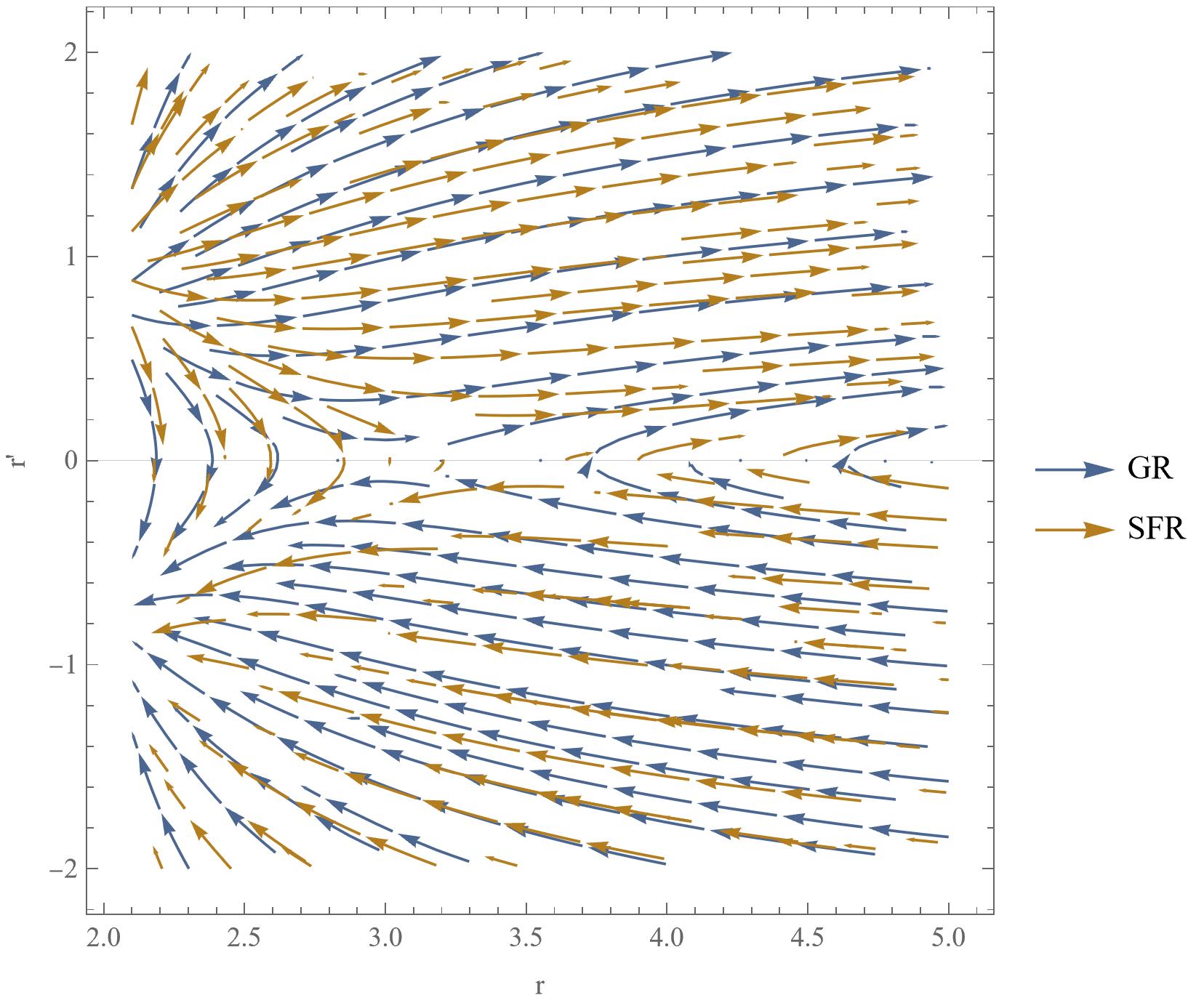}
\caption{This is a comparison between the radial phase portraits of the GR (blue) and SFR (yellow) models}
\label{phase-sfr-gr}
\end{figure}

\section{Deflection Angle}\label{sec: Deflection}
In this section, we will deal with the deflection angle of the SFR model and we will compare our findings with the corresponding ones of the GR model. In this consideration, we take into account photons that pass close to a central mass M.
%.\\[7pt]
From Eq.~\eqref{radial3} for photons, we put $\epsilon=0$ and we get:
\begin{equation}
\label{b1eq}
\frac{\dot{r}^{2}}{J^2}+\frac{f}{r^{2}}+
\frac{2\tilde{A_{0}}f^{1/2}}{Jb}=\frac{1}{b^2}
\end{equation}
where $b=J/\mathcal{E}_{R}$ is a composite
constant formed by the ratio of the angular momentum $J$
divided by the energy of the particle moving along the geodesic $\mathcal{E}_{R}$.

By using the Leibniz chain-rule $\dot{\phi}=\frac{d\phi}{d\tau}=\frac{d\phi}{dr}\frac{dr}{d\tau}=\frac{d\phi}{dr}\dot{r}$ with the relations \eqref{angular momentum} and \eqref{b1eq} we have:
\begin{equation}
\frac{\dot{r}^{2}}{\dot{\phi}^{2}}=r^{4}\left(\frac{1}{b^2}-\frac{f}{r^{2}}-
\frac{2\tilde{A_{0}}f^{1/2}}{Jb}\right)
\end{equation}
After some rearrangements we find:
\begin{equation}
\label{phideflect}
\frac{d\phi}{dr}=\frac{1}{r^2}\left(\frac{1}{b^2}-\frac{f}{r^{2}}-
\frac{2\tilde{A_{0}}f^{1/2}}{Jb}\right)^{-1/2}
\end{equation}
The deflection angle is calculated by the integration of \eqref{phideflect}:
\begin{equation}
\label{int1}
\Delta\phi_{SFR}=2\int_{r_1}^{\infty}\frac{dr}{r^2}\left[\frac{1}{b^2}-\frac{1}{r^{2}}\left(1-\frac{2GM}{r}\right)-
\frac{2\tilde{A_{0}}}{Jb}\left(1-\frac{2GM}{r}\right)^{1/2}\right]^{-1/2}
\end{equation}
where we have used $f=1-\frac{2GM}{r}$.\\
We perform a change of variables in the integral of  Eq.~\eqref{int1}:
\begin{equation}
\Delta\phi_{SFR}=2\int^{w_{1}}_{0}dw\left[1-w^{2}\left(1-\frac{2GM}{b}w\right)-2a\left(1-\frac{2GM}{b}w\right)^{1/2}\right]^{-1/2}\label{intwdeflection}
\end{equation}
where we have set $w=\frac{b}{r}$ and $a=\frac{\tilde{A_{0}}b}{J}$.\\
If we expand the integral in powers of $\frac{2GM}{b}$ and $a$ we find:
\begin{equation}
\Delta\phi_{SFR}\approx 2\int_{0}^{w_1}dw\frac{1+\frac{GM}{b}w}{\left[\left(1-2a\right)+\frac{2GM}{b}w-w^{2}\right]^{1/2}}
\label{intdeflection}
\end{equation}
where we have omitted  second order terms.\\ 
By evaluating the integral in Eq.~\eqref{intdeflection} we get (see Appendix 3):
\begin{equation}
\Delta\phi_{SFR}=\pi+\frac{4GM}{b}\frac{1-a}{\sqrt{1-2a}}    
\end{equation}
The deflection angle $\delta\phi_{SFR}$ can be found as:
\begin{align}
&\delta\phi_{SFR}=\Delta\phi_{SFR}-\pi\Rightarrow\nonumber\\[7pt]
&\delta\phi_{SFR}=\frac{4GM}{b}\frac{1-a}{\sqrt{1-2a}}    \label{SFRangle}    
\end{align}
If we expand Eq.~\eqref{SFRangle} in powers of $a=\frac{\tilde{A}_{0}b}{J}$ the deflection angle can be written as:
\begin{equation}    \delta\phi_{SFR}\approx \left(1+\frac{a^{2}}{2} \right)\frac{4GM}{b}
\end{equation}
The deflection angle $\delta\phi$ of GR \cite{Hartle2002, Carroll} 
is given by:
\begin{equation}
\label{angleGR}
\delta\phi_{GR}=\frac{4GM}{b}    
\end{equation}
Therefore, we notice that the deflection angle of SFR includes a small additional Randers contribution term $a$ which shows a small deviation from GR because $|\tilde{A_0}|\ll 1$.
We can see from Eq.~\eqref{SFRangle} that:
\begin{equation}
\lim_{\tilde{A_0}\rightarrow 0}\delta\phi_{SFR}=\delta\phi_{GR}
\end{equation}
The small difference of the deflection angle of the SFR model from the GR one can plausibly be attributed to the Lorentz violations \cite{Kostelecky:2011qz} or on the small amount of energy which is added to the gravitational potential of SFR. \\[7pt] 

S.~S.~Shapiro {\it et al.} \cite{Shapiro} have shown that the deflection angle for a light ray is:
\begin{equation}
 \theta\simeq\frac{(1+\gamma)GM}{c^{2}b}(1+cos\phi) \end{equation}
 where M is the mass of the Sun, G is the gravitational constant, $\phi$ is the angle between the source and the Sun for an observer on Earth and $\gamma$ is the PPN parameter that characterizes the contribution of space curvature to gravitational deflection and is estimated to be $\gamma_{Shap}=0.9998\pm 0.0004$. We compare this observational result of Shapiro {\it et al.} with  
 Eq.~(55) of our model in order to obtain some constraints for the parameter $\alpha$. If we do this we find:
\begin{equation}
    2 \left(1 + \frac{\alpha^{2}}{2}\right)\simeq 1 + \gamma_{Shap}
\end{equation}
 where we have taken $\phi\simeq 0$ for an infinitesimal neighbourhood of a fiducial geodesic.
 By solving for $\alpha$ we get:
 \begin{equation}
\alpha\simeq\pm\sqrt{-1+\gamma_{Shap}}   
 \end{equation}
which gives the values for $\alpha\simeq\pm 0.0141421$. This result provides an interesting relation between the deflection angle of a light ray in the SFR model and  the approach to the deflection angle used
in~\cite{Shapiro}.\\[7pt]

\textbf{\textit{Remark 3}}: By considering the following relation, we can connect the geometrical concept of the curvature $\kappa_{\phi}=\frac{d\phi}{d\tau}$ of a path with the deflection angle $\delta\phi$ in the following way:    
\begin{align}
 \dot{\phi}=\frac{d\phi}{d\tau}&=\frac{d\phi}{dr}\frac{dr}{d\tau}=\frac{d\phi}{dr}\dot{r}\Rightarrow\\
 \label{chainrule}
 &\kappa_{\phi}=\frac{d\phi}{dr}\dot{r}\Rightarrow\\
 &\delta\phi=\int\frac{\kappa_{\phi}}{\dot r} dr
\end{align}
This form of curvature can be called \textit{deflection curvature}.

\section{Conclusions \& Future Challenges}\label{sec: Conclusions}
In this article, we investigated the analytic form of the geodesics of the model SFR which was introduced in previous works \cite{Triantafyllopoulos:2020vkx,Kapsabelis:2021dpb}. A dynamical analysis was presented based on the energy and angular momentum of a particle along of geodesics (null or timelike) of 
the SFR spacetime. Comparisons  between the SFR and GR 
were provided. We found that there is a  small deviation from 
the GR model which is due to the dynamical term $A_{\gamma }(x)$.
%(rel.~\ref{RandersL}).
%In the section IV 
We also formulated and studied an effective potential of our model and we compared the one of the SFR case once again 
with the effective potential of GR attributing
the small but discernible differences to the specific
structure of (and perturbation incorporated within) 
the SFR spacetime. The relevant differences in the
trajectories were illustrated both in the evolution 
over the time-variable $\tau$ and in the $(x,y)$ plane.
%the small difference can be attributed in the extra term from the rel.(\ref{VeffSFReq}) which is produced by the structure of SFR spacetime (rel.\ref{RandersL}).\\
In addition, we calculated the deflection angle 
%($\epsilon=0$) 
for the SFR spacetime and we compared with the corresponding one of GR. The result is a small difference of the
SFR model from GR, it is possibly caused by Lorentz violations or by the small amount of energy which is added to the gravitational potential of SFR spacetime.

In addition, we found an interesting relation between the observational data of~\cite{Shapiro} and the SFR model's prediction for the deflection angle of a null geodesic. Finally, we presented an application which relates the covector field $A_\gamma(x)$ of the SFR model with the Newtonian gravitational potential.

It is important to note that this work opens a number of
interesting directions of further study for the future. 
On the one hand, the traditional assumption of $\theta=\pi/2$
made over here is clearly a restrictive one that
simplifies the equations of motion automatically satisfying
the dynamics for the angular variable $\theta$ with the
latter being at steady state. However, more generally,
one can straightforwardly envision scenarios
where this condition is no longer satisfied.
It is then of interest to explore 
if one
starts in the vicinity of $\pi/2$ whether one stays
in that neighborhood or perhaps if one deviates
away from this steady state and how the associated
dynamics of the full 4-degree-of-freedom space is accordingly
explored.
Another aspect that is also worth further exploring
is that of the small amplitude covector deviation
from the General Relativity standard model. Here,
we have limited our considerations to the realm
of associated small amplitude perturbations (where
leading order expansions of the field would suffice).
However, it would also be of interest to explore the situation
when one gradually deviates from the realm of this
approximation as well.
{In addition, applications of geodesics of the SFR model can be pursued for
more concrete cosmological studies such as, e.g., for the case of the S2 stars orbiting the black hole in Sagittarius A* in which the geodesics of the star are perturbed from the classical Keplerian orbits because of the distribution of stellar remnants.
}  Indeed, our hope is that this work may 
pave the way towards testing the Schwarzschild-Finsler-Randers gravitational model which incorporates features going beyond the standard Riemannian geometry of spacetime.
In this vein, some of the above topics are presently
under consideration and associated results will
be presented in future publications.

\section{Acknowledgements}
The authors would like to thank Prof. Gary Gibbons for his valuable suggestions.
\appendix

\section{Distinguished connection on $TM$.}\label{sec:d-connection}

	In this work, we consider a distinguished connection ($d-$connection) $ {D} $ on $TM$ \cite{Miron:1994nvt,Vacaru:2005ht}. This is a linear connection with coefficients $\{\Gamma^A_{BC}\} = \{L^\mu_{\nu\kappa}, L^\alpha_{\beta\kappa}, C^\mu_{\nu\gamma}, C^\alpha_{\beta\gamma} \} $ which preserves by parallelism the horizontal and vertical distributions:
	\begin{align}
	{D_{\delta_\kappa}\delta_\nu = L^\mu_{\nu\kappa}(x,y)\delta_\mu} \quad &,\quad D_{\pdot{\gamma}}\delta_\nu = C^\mu_{\nu\gamma}(x,y)\delta_\mu \label{D delta nu} \lin
	{D_{\delta_\kappa}\pdot{\beta} = L^\alpha_{\beta\kappa}(x,y)\pdot{\alpha}} \quad &, \quad D_{\pdot{\gamma}}\pdot{\beta} = C^\alpha_{\beta\gamma}(x,y)\pdot{\alpha} \label{D partial b}
	\end{align}
	From the above conditions, the definitions for partial covariant differentiation follow immediately, e.g. for $X \in TTM$ the expression for the covariant h-derivative is:
	\begin{equation}
	X^A_{|\nu} \equiv D_\nu\,X^A \equiv \delta_\nu X^A + L^A_{B\nu}X^B \label{vector h-covariant}
	\end{equation}
	and for the covariant v-derivative:
	\begin{equation}
	X^A|_\beta \equiv D_\beta\,X^A \equiv \dot{\partial}_\beta X^A + C^A_{B\beta}X^B \label{vector v-covariant}
	\end{equation}
	The $d-$connection is metric-compatible when we have:
	\begin{equation}
	D_\kappa\, g_{\mu\nu} = 0, \quad D_\kappa\, v_{\alpha\beta} = 0, \quad D_\gamma\, g_{\mu\nu} = 0, \quad D_\gamma\, v_{\alpha\beta} = 0
	\end{equation}
A $d-$connection can be uniquely defined when the following conditions are satisfied:
\begin{itemize}
	\item The $d-$connection is metric compatible
	\item Coefficients $L^\mu_{\nu\kappa}, L^\alpha_{\beta\kappa}, C^\mu_{\nu\gamma}, C^\alpha_{\beta\gamma} $ depend solely on the quantities $g_{\mu\nu}$, $v_{\alpha\beta}$ and $N^\alpha_\mu$
	\item Coefficients $L^\mu_{\kappa\nu}$ and $ C^\alpha_{\beta\gamma} $ are symmetric on the lower indices, i.e.  $L^\mu_{[\kappa\nu]} = C^\alpha_{[\beta\gamma]} = 0$
\end{itemize}
	We use the symbol $\mathcal D$ instead of $D$ for a connection satisfying these conditions. We call $\mathcal D$ a canonical and distinguished $d-$connection.
	The coefficients of this connection are
	%can be found in \cite{miron-watanabe-ikeda 1987}:
	\begin{align}
	L^\mu_{\nu\kappa} & = \frac{1}{2}g^{\mu\rho}\left(\delta_kg_{\rho\nu} + \delta_\nu g_{\rho\kappa} - \delta_\rho g_{\nu\kappa}\right) \label{metric d-connection 1_a}  \\
	L^\alpha_{\beta\kappa} & = \dot{\partial}_\beta N^\alpha_\kappa + \frac{1}{2}v^{\alpha\gamma}\left(\delta_\kappa v_{\beta\gamma} - v_{\delta\gamma}\,\dot{\partial}_\beta N^\delta_\kappa - v_{\beta\delta}\,\dot{\partial}_\gamma N^\delta_\kappa\right) \label{metric d-connection 2_a}  \\
	C^\mu_{\nu\gamma} & = \frac{1}{2}g^{\mu\rho}\dot{\partial}_\gamma g_{\rho\nu} \label{metric d-connection 3_a} \\
	C^\alpha_{\beta\gamma} & = \frac{1}{2}v^{\alpha\delta}\left(\dot{\partial}_\gamma v_{\delta\beta} + \dot{\partial}_\beta v_{\delta\gamma} - \dot{\partial}_\delta v_{\beta\gamma}\right) \label{metric d-connection 4_a}
	\end{align}
	
	 Curvatures and torsions on $TM$ are defined by the linear maps:
	\begin{equation}
	\mathcal{R}(X,Y)Z = [\mathcal{D}_X,\mathcal{D}_Y]Z - \mathcal{D}_{[X,Y]}Z \label{Riemann tensor TM}
	\end{equation}
	and
	\begin{equation}
	\mathcal{T}(X,Y) = \mathcal{D}_XY - \mathcal{D}_YX - [X,Y] \label{torsion TM}
	\end{equation}
	where $X,Y,Z \in TTM$.
	We use the following definitions for the curvature components \cite{Miron:1994nvt,Vacaru:2005ht}:
	\begin{align}
	\mathcal{R}(\delta_\lambda,\delta_\kappa)\delta_\nu = R^\mu_{\nu\kappa\lambda}\delta_\mu \label{R curvature components} \lin
	\mathcal{R}(\delta_\lambda,\delta_\kappa)\pdot{\beta} = R^\alpha_{\beta\kappa\lambda}\pdot{\alpha}\lin
	\mathcal{R}(\pdot{\gamma},\delta_\kappa)\delta_\nu = P^\mu_{\nu\kappa\gamma}\delta_\mu \lin
	\mathcal{R}(\pdot{\gamma},\delta_\kappa)\pdot{\beta} = P^\alpha_{\beta\kappa\gamma}\pdot{\alpha}\lin
	\mathcal{R}(\pdot{\delta},\pdot{\gamma})\delta_\nu = S^\mu_{\nu\gamma\delta}\delta_\mu\lin
	\mathcal{R}(\pdot{\delta},\pdot{\gamma})\pdot{\beta} = S^\alpha_{\beta\gamma\delta}\dot{\partial}_\alpha \label{S curvature components}
	\end{align}
	In addition, we use the following definitions for the torsion components:
	\begin{align}
	\mathcal{T}(\delta_\kappa,\delta_\nu) = & \mathcal{T}^\mu_{\nu\kappa}\delta_\mu + \mathcal{T}^\alpha_{\nu\kappa}\pdot{\alpha} \label{torsion components 1} \lin
	\mathcal{T}(\pdot{\beta},\delta_\nu) = & \mathcal{T}^\mu_{\nu\beta}\delta_\mu + \mathcal{T}^\alpha_{\nu\beta}\pdot{\alpha} \label{torsion components 2} \lin
	\mathcal{T}(\pdot{\gamma},\pdot{\beta}) = & \mathcal{T}^\mu_{\beta\gamma}\delta_\mu + \mathcal{T}^\alpha_{\beta\gamma}\pdot{\alpha} \label{torsion components 3}
	\end{align}
	From \eqref{R curvature components}, the h-curvature tensor of the $d-$connection in the adapted basis and the corresponding h-Ricci tensor read:
	\begin{align}
	& R^\mu_{\nu\kappa\lambda} = \delta_\lambda L^\mu_{\nu\kappa} - \delta_\kappa L^\mu_{\nu\lambda} + L^\rho_{\nu\kappa}L^\mu_{\rho\lambda} - L^\rho_{\nu\lambda}L^\mu_{\rho\kappa} + C^\mu_{\nu\alpha}\Omega^\alpha_{\kappa\lambda} \label{R coefficients 1}\lin
	& R_{\mu\nu} = R^\kappa_{\mu\nu\kappa} =  \delta_\kappa L^\kappa_{\mu\nu} - \delta_\nu L^\kappa_{\mu\kappa} + L^\rho_{\mu\nu}L^\kappa_{\rho\kappa} - L^\rho_{\mu\kappa}L^\kappa_{\rho\nu} + C^\kappa_{\mu\alpha}\Omega^\alpha_{\nu\kappa} \label{d-ricci 1}
	\end{align}
	From \eqref{S curvature components}, the v-curvature tensor of the $d-$connection in the adapted basis and the corresponding v-Ricci tensor are:
	\begin{align}
	S^\alpha_{\beta\gamma\delta} & = \pdot{\delta} C^\alpha_{\beta\gamma} - \pdot{\gamma}C^\alpha_{\beta\delta} + C^\epsilon_{\beta\gamma}C^\alpha_{\epsilon\delta} - C^\epsilon_{\beta\delta}C^\alpha_{\epsilon\gamma} \label{S coefficients 2} \lin
	S_{\alpha\beta} & = S^\gamma_{\alpha\beta\gamma} = \pdot{\gamma}C^\gamma_{\alpha\beta} - \pdot{\beta}C^\gamma_{\alpha\gamma} + C^\epsilon_{\alpha\beta}C^\gamma_{\epsilon\gamma} - C^\epsilon_{\alpha\gamma}C^\gamma_{\epsilon\beta} \label{d-ricci 4}
	\end{align}
	The generalized Ricci scalar curvature in the adapted basis is:
	\begin{equation}
	\R = g^{\mu\nu}R_{\mu\nu} + v^{\alpha\beta}S_{\alpha\beta} = R+S \label{bundle ricci curvature}
	\end{equation}
	where
	\begin{align}
	R=g^{\mu\nu}R_{\mu\nu} \quad,\quad
	S=v^{\alpha\beta}S_{\alpha\beta} \label{hv ricci scalar}
	\end{align}
	
\section{Field equations of the model.}\label{sec:field_eqs}
	
	A Hilbert-like action on $TM$ can be defined as
	\begin{equation}\label{Hilbert like action}
	K = \int_{\mathcal N} d^8\mathcal U \sqrt{|\Gd|}\, \R + 2\kappa \int_{\mathcal N} d^8\mathcal U \sqrt{|\Gd|}\,\mathcal L_M
	\end{equation}
		for some closed subspace $\mathcal N\subset TM$, where $|\Gd|$ is the absolute value of the metric determinant, $\mathcal L_M$ is the Lagrangian of the matter fields, $\kappa$ is a constant and
	\begin{equation}
	d^8\mathcal U = \de x^0 \wedge \ldots \wedge\de x^3 \wedge \de y^4 \wedge \ldots \wedge \de y^7
	\end{equation}
	Variation with respect to $g_{\mu\nu}$, $v_{\alpha\beta}$ and $N^\alpha_\kappa$ leads to the following field equations \cite{Triantafyllopoulos:2020ogl}:
	\begin{align}
		& \overline R_{\mu\nu} - \frac{1}{2}({R}+{S})\,{g_{\mu\nu}}  + \left(\delta^{(\lambda}_\nu\delta^{\kappa)}_\mu - g^{\kappa\lambda}g_{\mu\nu} \right)\left(\mathcal D_\kappa\mathcal T^\beta_{\lambda\beta} - \mathcal T^\gamma_{\kappa\gamma}\mathcal T^\beta_{\lambda\beta}\right)  = T_{\mu\nu}  \label{feq1}\\
		& S_{\alpha\beta} - \frac{1}{2}({R}+{S})\,{v_{\alpha\beta}} + \left(v^{\gamma\delta}v_{\alpha\beta} - \delta^{(\gamma}_\alpha\delta^{\delta)}_\beta \right)\left(\mathcal D_\gamma C^\mu_{\mu\delta} - C^\nu_{\nu\gamma}C^\mu_{\mu\delta} \right) = Y_{\alpha\beta} \label{feq2}\\
		& g^{\mu[\kappa}\pdot{\alpha}L^{\nu]}_{\mu\nu} +  2 \mathcal T^\beta_{\mu\beta}g^{\mu[\kappa}C^{\lambda]}_{\lambda\alpha} = \mathcal Z^\kappa_\alpha \label{feq3}
	\end{align}
	 where
	\begin{equation}\label{torsion}
		\mathcal{T}_{\nu\beta}^{\alpha} = \pdot{\beta} N_{\nu}^{\alpha} - L_{\beta\nu}^{\alpha}
	\end{equation}
	are torsion components, where $L_{\beta\nu}^{\alpha}$ is defined in \eqref{metric d-connection 2}. From the form of \eqref{bundle metric} it follows that $\sqrt{|\Gd|} = \sqrt{-g}\sqrt{-v}$, with $g, v$ the determinants of the metrics $g_{\mu\nu}, v_{\alpha\beta}$ respectively.
\section{Calculation of the deflection angle}
We begin the calculation from Eq.~\eqref{intwdeflection}
\begin{equation}
\Delta\phi_{SFR}=2\int^{w_{1}}_{0}dw\left[1-w^{2}\left(1-\frac{2GM}{b}w\right)-2a\left(1-\frac{2GM}{b}w\right)^{1/2}\right]^{-1/2}
\end{equation}
\begin{align}
&\Delta\phi_{SFR}=2\int_{0}^{w_1}dw\left(1-\frac{2GM}{b}w\right)^{-1/2}\left[\left(1-\frac{2GM}{b}w\right)^{-1}-w^{2}-
2a\left(1-\frac{2GM}{b}w\right)^{-1/2}\right]^{-1/2}\nonumber\Rightarrow\\[7pt]
&\Delta\phi_{SFR}\approx 2\int_{0}^{w_1}dw\left(1+\frac{GM}{b}w\right)\left[\left(1+\frac{2GM}{b}w\right)-w^{2}-
2a\right]^{-1/2}\nonumber\Rightarrow\\[7pt]
&\Delta\phi_{SFR}\approx 2\int_{0}^{w_1}dw\frac{1+\frac{GM}{b}w}{\left[\left(1+\frac{2GM}{b}w\right)-w^{2}-
2a\right]^{1/2}}\nonumber\Rightarrow\\[7pt]
&\Delta\phi_{SFR}\approx  2\int_{0}^{w_1}dw\frac{1+\frac{GM}{b}w}{\left[\left(1-2a\right)+\frac{2GM}{b}w-w^{2}\right]^{1/2}}
\label{appint}
\end{align}
In order to find $w_{1}$ we solve the following equation from the denominator:
\begin{equation}
\left(1-2a\right)+\frac{2GM}{b}w-w^{2}=0
\end{equation}
and we get:
\begin{equation}
w_{1}=\frac{GM}{b}+\sqrt{\left(\frac{GM}{b}\right)^{2}+(1-2a)}
\end{equation}
which is the positive root of the denominator.
The solution for the integral in Eq.~\eqref{appint} is:
\begin{equation}
\Delta\phi_{SFR}=\pi +\frac{1}{\sqrt{1-2a}}\frac{2GM}{b}+\frac{2GM}{b}\sqrt{1-2a}
\end{equation}
where we omit terms $O((\frac{2GM}{b})^{2})$, given their
smallness.
Hence, we find:
\begin{equation}
\Delta\phi_{SFR}=\pi+\frac{4GM}{b}\frac{1-a}{\sqrt{1-2a}}    
\end{equation}

\end{document}